\documentclass[11pt]{article}
\usepackage[margin=1.0in]{geometry}
\usepackage{graphicx}
\usepackage{epsfig}
\usepackage{amssymb,amsfonts,amsmath,amsthm}
\usepackage{mathrsfs}
\usepackage{epstopdf}
\usepackage{hyperref,color}
\usepackage{natbib}
\usepackage{setspace}
\usepackage[multiple]{footmisc}

\newtheorem{defi}{Definition}

\newtheorem{theorem}{Theorem}

\newcommand{\pl}{\ell_{\mbox{\scriptsize{Pl}}}}

\begin{document}

\title{Raiders of the lost spacetime}
\author{Christian W\"uthrich\thanks{I thank audiences at the University of London's Institute of Philosophy and at the spacetime workshop at the Bergisch University of Wuppertal. I am also grateful to my fellow Young Guns of General Relativity, Erik Curiel, John Manchak, Chris Smeenk, and Jim Weatherall for valuable discussions and comments, and to Francesca Vidotto for correspondence. Finally, I owe thanks to two perceptive referees for their comments. Work on this project has been supported in part by a Collaborative Research Fellowship by the American Council of Learned Societies, by a UC President's Fellowship in the Humanities, and by the Division for Arts and Humanities at the University of California, San Diego. Some of the material in this essay has its origin in \citet{wut06}.}}
\date{For Dennis Lehmkuhl (ed.), {\em Towards a Theory of Spacetime Theories}. Birkh\"auser.}
\maketitle

\begin{abstract}\noindent
Spacetime as we know and love it is lost in most approaches to quantum gravity. For many of these approaches, as inchoate and incomplete as they may be, one of the main challenges is to relate what they take to be the fundamental non-spatiotemporal structure of the world back to the classical spacetime of GR. The present essay investigates how spacetime is lost and how it may be regained in one major approach to quantum gravity, loop quantum gravity.

\vspace{3mm}\noindent
\end{abstract}

\noindent
Many approaches to quantum gravity (QG) suggest or imply that space and time do not exist at the most fundamental ontological level, at least not in anything like their usual form. Thus deprived of their former status as part of the fundamental furniture of the world, together, perhaps, with quarks and leptons, they merely `emerge' from the deeper physics that does not rely on, or even permit, their (fundamental) existence, rather like tables and chairs. The extent to which the fundamental structures described by competing approaches to QG diverge from relativistic spacetimes varies, along different dimensions \citep{hugwut13b}. That modern physics puts time under pressure is widely accepted. One can read the history of modern physics from the advent of relativity theory to the present day as a continuing peeling away of the structure that time was initially believed to exemplify \citep*[\S2.1]{hugeal13}. But at least in some approaches, spacetime as a whole comes under siege. This may occur in the relatively mild sense that the fundamental structure turns out to be discrete; or it may be discrete and non-local, as it happens in loop quantum gravity (LQG); or the reality of some dimensions of space is questionable altogether, as it is in theories with certain dualities; or it may exhibit non-commutativity among different dimensions, obliterating the usual geometric understanding that we routinely have of spacetime. 

Just how radical the departure from the spacetime we know and love is remains to be seen, but it is likely to have profound implications. For instance, it may render some of our cherished philosophical theories not just of space and time, but also of persistence, causation, laws of nature, and modality obsolete, or at least in need of revision \citep{wut12b}. But this paper will be concerned with the consequences for the physics, rather than the metaphysics. Two urgent, and related, issues arise. First, one might worry that if it is a necessary condition for an empirical science that we can at least in principle measure or observe something {\em at some location at some time}. The italicized locution, in turn, seems to presuppose the existence of space and time. If that existence is now denied in quantum theories of gravity, one might then fear that these theories bid adieu to empirical science altogether. It thus becomes paramount for advocates of these theories to show that the latter only threaten the {\em fundamentality}, but not the {\em existence} of space and time. To discharge this task means to show how relativistic spacetimes re-emerge and how measurable quantities arise from the fundamental structure as postulated by the theory at stake. 

This first issue is closely related to a second problem: a novel theory can supplant an incumbent theory only if it recreates at least most of the empirical success of the old theory. The way in which this requirement is typically met in physics is by showing how the newer theory offers a more general framework than the older one, and that therefore the older is a special case of the newer, which can be regained, or at least mocked in formally suggestive ways, in some limit or to some approximation. For instance, it was important to Albert Einstein to be able to show that one obtains from general relativity (GR), in a weak-field limit, a theory which returns essentially all the same empirical results in the appropriate regime as Newtonian gravitational theory. This recovery mattered because the Newtonian theory garnered impressive empirical successes over the more than two centuries preceding Einstein's formulation of GR. For the very same reason, present-day quantum theories of gravity must eventually prove that they relate, in physically salient ways, to the classical GR that the last century of observations has found to be so accurate.\footnote{Given this formidable success of the classical theory, one might wonder why we need a quantum theory of gravity at all. There are good reasons to think that we do, but they do not fully align with the standard lore one finds in the physics literature \citep[\S1]{wut13a}.} In fact, given the complete absence of direct empirical access to the quantum-gravitational regime, establishing this link with `old' physics arguably constitutes the single most important constraint on theorizing in the quantum-gravitational realm.

Consequently, in theories of lost spacetime, relativistic spacetimes must be regained from the fundamental structure in order to discharge the tasks of securing both the theory's empirical coherence and its account of why the theory it seeks to supplant was as successful as it was. It is the goal of this essay to show just how spacetime vanishes and how it might be seen to re-emerge in one important approach to quantum gravity, LQG. Since the emergence of spacetime from a non-spatiotemporal structure is often thought to be impossible, establishing the mere {\em possibility} of such emergence assumes vital importance.\footnote{For a very recent critical view, see e.g.~\citet{lamesf13}.}

The next section, Section \ref{sec:probtime}, explicates how time, rather than spacetime, disappears in a class of approaches to QG, the so-called `canonical' theories. Canonical QG casts GR in a particular way, and the section will show how time and change vanish already at the level of GR so cast. Section \ref{sec:dissolve} then investigates the fundamental structures as they are described by LQG and discusses the two main ways in which they differ from relativistic spacetimes, viz.\ in their discreteness and their non-locality. The following section, Section \ref{sec:nonem}, starts to clear the path for the re-emergence of relativistic spacetime by arguing how the emergence relation should {\em not} be construed in the present case. Specifically, it argues against a non-reductive understanding of emergence and an attempt to cash out the relation between the structures in terms of unitary equivalence as both inadequate to the task at hand. Next, Section \ref{sec:reemerge} sketches a way in which the relationship between fundamental spin networks and relativistic spacetimes might be worked out and tries to understand what it would generally take to relate them. Section \ref{sec:conc} offers brief conclusions.

\section{The problem of time in canonical general relativity}\label{sec:probtime}

Casting GR as a Hamiltonian system with constraints has many advantages, as John \citet{ear03} affirmed: it gives the vague talk about `local' and `global' transformations a more tangible meaning, it explains how the fibre bundle formalism arises in the cases it does, it has a sufficiently broad scope to relate GR to Yang-Mills gauge theories, it offers a formalization of the gauge concept, and it connects to  foundational issues such as the nature of observables and the status of determinism in GR and in gauge theories. Moreover, the Hamiltonian formulation affords a natural affinity to the initial value problem in GR.\footnote{Cf.\ \citet[Appendix E.2]{walgr}. A {\em locus classicus} for the Cauchy problem in GR is \citet{choyor80}; a more recent survey article is \citet{friren00}.} The real gain of a Hamiltonian formulation, however, arises when one tries to quantize the classical theory. Typically, prescriptions to find a quantum theory from a classical theory require either a Lagrangian (e.g.\ for the path integral method) or a Hamiltonian (e.g.\ for canonical quantization) formulation of the theory. LQG relies on a canonical quantization procedure and thus uses a Hamiltonian formulation of GR as a starting point.\footnote{A useful introduction to the Lagrangian and the Hamiltonian formulation of GR is given in \citet[Appendix E]{walgr}. Wald's textbook of 1984 only deals with the ADM version of Hamiltonian GR and, as time travel was not yet invented in 1984, does not treat Ashtekar's version, pioneered in 1986.}\footnote{Of course, for most cases we care about, Hamiltonian theories afford a corresponding equivalent Lagrangian theory, and vice versa. Currently, a debate rages in philosophy of physics over which of the two, if any, is more fundamental or more perspicuous. Nothing I say here should be taken to entail a stance in that debate.}

However, forcing GR, to use the words of Tim \citet[9]{mau02}, ``into the Procrustean bed of the Hamiltonian formalism'' also comes, as conveyed by the quote, at a cost. The cost arises from the fact that the Hamiltonian formalism tends to construe the physical systems it describes as spatially extended three-dimensional objects evolving over an external time, and this is no different for the Hamiltonian formulation of GR.\footnote{There are, of course, purely internal degrees of freedom of particles, such as classical spin, which admit of a Hamiltonian treatment without the system necessarily being extended in space. Now, even a point particle with internal degrees of freedom is at least a physical system {\em in} space, and it certainly also evolves over external time.} Recasting GR in a Hamiltonian formalism thus reinterprets the four-dimensional spacetimes of standard GR as three-dimensional `spaces' which evolve in a fiducial `time' according to the dynamics governed by Hamilton's equation. Pulling space and time asunder in this way, of course, contravenes the received view of what many take to be the deepest insight of relativity, viz.\ that no separation of the fundamental {\em spacetime} into space and time can in any physically relevant way be privileged. This blatant violation of four-dimensionalism, of course, gets mathematically mended in the formalism through the imposition of constraints. But we are getting ahead of ourselves. What this brief paragraph should suggest is that having a philosophically closer look at the dynamics of this reformulation of classical GR is worth our while.\footnote{In connection with what follows, Chapter 1 of \citet{hentei} is recommended reading. For a less formal and hence more accessible treatment of the problem of time, cf.\ \citet[\S2]{hugeal13} and references therein. Cf.\ also Kiefer's contribution to this collection.}

A spacetime is an ordered pair $\langle \mathcal{M}, g_{ab}\rangle$ consisting of a four-dimensional pseudo-Riemannian manifold $\mathcal{M}$ and a metric tensor field $g_{ab}$ defined on $\mathcal{M}$. Starting out from the Einstein-Hilbert action $S[g_{ab}]$ for gravity without matter,
\begin{equation}\label{eq:hilact}
S[g_{ab}] = \frac{1}{16\pi G} \int_\mathcal{M} d^4x \sqrt{-g} R,
\end{equation}
where $G$ is Newton's gravitational constant, $g$ the determinant of the metric tensor $g_{ab}$, and $R$ the Ricci scalar,  one can gain a Lagrangian formulation of GR with the dynamical Euler-Lagrange equations in terms of a Lagrangian function $L(q, \dot{q})$ of generalized coordinates $q$ and the generalized velocities $\dot{q}$. The Lagrange function is essentially the integrand in the action integral (\ref{eq:hilact}) integrated over the three spatial dimensions. This action leads to the (vacuum) field equations of GR if one varies (\ref{eq:hilact}) with respect to the metric $g_{ab}$. Thus, Einstein's vacuum field equations can be recognized as the equations of motion of the Lagrangian formulation of GR, i.e.\ as the Euler-Lagrange equations. They are second-order differential equations. The solutions to the Euler-Lagrange equations will be uniquely determined by $q,\dot{q}$ just in case the so-called `Hessian' matrix $\partial^2 L(q,\dot{q})/\partial \dot{q}^{n'}\partial \dot{q}^n$ of $L(q, \dot{q})$, where $n$ labels the degrees of freedom, is invertible. This is the case if and only if its determinant, confusingly sometimes also called `Hessian', does not vanish.  In case the determinant of the Hessian vanishes, which means the Hessian is `singular', the accelerations $\ddot{q}$ will not be uniquely determined by the positions and the velocities and the solutions to the Euler-Lagrange equations are not only not unique in $q$ and $\dot{q}$, but also contain arbitrary functions of time. Thus, the impossibility of inverting $\partial^2 L(q,\dot{q})/\partial \dot{q}^{n'}\partial \dot{q}^n$ is an indication of gauge freedom. How such gauge freedom arises in constrained Hamiltonian systems is the topic of the next subsection, \S\ref{ssec:hamcon}, followed by an analysis in \S\ref{ssec:hamgr} of how this lesson carries over into the context of Hamiltonian GR and leads to the problem of time. 

\subsection{Hamiltonian systems with constraints}\label{ssec:hamcon}

Finding a Hamiltonian formulation amounts to putting the Euler-Lagrange equations in the form of Hamiltonian equations of motion, $\dot{q} = \partial H/\partial p$ and $\dot{p} = \partial H/\partial q$, which are of first order. This can be achieved by the introduction of canonical momenta via
\begin{equation}\label{eq:canon}
p_n = \frac{\partial L}{\partial \dot{q}^n},
\end{equation}
where $n=1,...,N$, $N$ being the number of degrees of freedom of the system at stake. These momenta are not all independent when we are faced with a system exhibiting gauge freedom---i.e.\ just in case the Hessian is singular. These dependencies get articulated in constraint equations
\begin{equation}\label{eq:primary}
\phi_m (q,p) =0, \;\; m=1,...,M,
\end{equation}
where $M$ is the number of dependencies. The relations (\ref{eq:primary}) between $q$ and $p$ are called {\em primary constraints} and define a submanifold smoothly embedded in phase space called the {\em primary constraint surface}. The phase space $\Gamma$ is defined as the space of solutions of the equations of motion. Assuming that all equations (\ref{eq:primary}) are linearly independent, which may not be the case, this submanifold will be of dimension $2N-M$. Equations (\ref{eq:primary}) imply that the transformation map between the Lagrangian phase space $\Gamma(q,\dot{q})$ and the Hamiltonian phase space $\Gamma(q,p)$ is onto but not one-to-one. Equations (\ref{eq:canon}) define a mapping from a $2N$-dimensional manifold of the $q$'s an $\dot{q}$'s to the $(2N-M)$-dimensional manifold defined by (\ref{eq:primary}). In order to render the transformation bijective and thus invertible, the introduction of extra parameters---`gauge fluff'---is required.\footnote{For more details on how the constraints arise in some Hamiltonian systems, see \citet[Ch.~1]{hentei}. My exposition largely follows this reference.}

Next, one introduces a Hamiltonian $H$ as a function of position and momentum variables as
\begin{equation}\label{eq:legendre}
H (q,p) = \dot{q}^n p_n - L (q,\dot{q}).
\end{equation}
This canonical Hamiltonian is uniquely defined only on the primary constraint surface but can arbitrarily be extended to the rest of phase space. The `Legendre transformation' defined by (\ref{eq:canon}) turns out to be invertible just in case $\det(\partial^2 L/ \partial \dot{q}^{n'} \partial \dot{q}^n)\neq 0$. Should the determinant of the Hessian vanish, as above, one can add extra variables $u^m$ and thus render the Legendre transformation invertible. In this case, the Hamiltonian equations corresponding to the Euler-Lagrange equations become
\begin{eqnarray}
& & \dot{q}^n = \frac{\partial H}{\partial p_n} + u^m\frac{\partial \phi_m}{\partial p_n},\nonumber \\
& & \dot{p}_n = -\frac{\partial H}{\partial q^n} - u^m\frac{\partial \phi_m}{\partial q^n},\nonumber \\
& & \phi_m (q,p) = 0.\nonumber
\end{eqnarray}
These Hamilton equations lead via arbitrary variations $\delta q^n, \delta p_n, \delta u^m$ (except for the boundary conditions $\delta q^n (t_1) = \delta q^n(t_2) =0$ and that they must conserve $H$) to the Hamiltonian equations of motion for arbitrary functions $F(q,p)$ of the canonical variables
\begin{equation}\label{eq:hammotion}
\dot{F} = \{F,H\} + u^m \{F, \phi_m\},
\end{equation}
where $\{,\}$ is the usual Poisson bracket
\[
\{F,G\} := \frac{\partial F}{\partial q^i} \frac{\partial G}{\partial p_i} - \frac{\partial F}{\partial p_i} \frac{\partial G}{\partial q^i}.
\]

Consistency requires that the primary constraints $\phi_m$ be preserved over time, i.e.\ that $\dot{\phi}_m=0$. As primary constraints are phase space functions, equation (\ref{eq:hammotion}) then implies
\begin{equation}\label{eq:secondary}
\{\phi_m, H\} + u^{m'} \{\phi_m, \phi_{m'}\} =0.
\end{equation}
This equation has one of two possible forms: either it embodies a relation only between the $q$'s and $p$'s, without any $u^m$, or it results in a relation including $u^m$. In the latter case, we just end up with a restriction on $u^m$. In the former case, however, (\ref{eq:secondary}) leads to additional constraints, called {\em secondary constraints}, on the canonical variables and thus on the physically relevant region of the phase space. These secondary constraints must also fulfill the consistency requirement of being preserved over time, which leads to new equations of the type (\ref{eq:secondary}), which again are either restrictions on the $u^m$ or constraints on the canonical variables, etc. Once the process is finished, and we have all secondary constraints\footnote{They are {\em not} referred to as tertiary, quaternary etc.\ constraints, but only collectively as `secondary' constraints.}, denoted by $\phi_k = 0$ with $k=M+1,...,M+K$, all constraints can be rewritten as $\phi_j =0$ with $j=1,..., M + K =: J$. The full set of constraints $\phi_j=0$ defines a `subsubmanifold' in the phase space $\Gamma$, i.e.\ a submanifold of the primary constraint surface $\phi_m=0$, called the {\em constraint surface} $\mathcal{C}$. The relevant difference between primary and secondary constraints is that primary constraints are direct consequences of equation (\ref{eq:canon}), whereas the secondary constraints only arise once the equations of motion (\ref{eq:hammotion}) are given.

Any two functions $F$ and $G$ in phase space that coincide on the constraint surface are said to be {\em weakly equal}, symbolically $F\approx G$. In case they agree throughout the entire phase space, their equality is considered {\em strong}, expressed as usual as $F=G$. Above, I have introduced the qualification of constraints as primary. However, there is a more important classification of constraints into first-class and second-class constraints, defined as follows:
\begin{defi}[First-class constraints]\label{def:firstclass}
A function $F(q,p)$ is termed {\em first class} if and only if its Poisson bracket with every constraint vanishes weakly,
\begin{equation}
\{F,\phi_j\} \approx 0, \;\; j=1,...,J.
\end{equation}
If that first-class function is a constraint itself, then we call it a {\em first-class constraint}. A function in phase space is called {\em second class} just in case it is not first class.
\end{defi}
The property of being first class is preserved under the Poisson bracket, i.e.\ the Poisson bracket of two first-class functions is first class again. 

The fact that arbitrary functions $u^m$ enter the Hamilton equations (or, equivalently, the Hamiltonian equations of motion) implies that a physical state is uniquely determined by a pair $(q,p)$, i.e.\ by a point in (Hamiltonian) phase space $\Gamma(q,p)$, but not vice versa. In other words, these arbitrary functions encode the gauge freedom which arises for systems with a singular Hessian. It can be shown that a dynamical variable $F$, i.e.\ a function on $\Gamma$, differs in value from time $t_1$ to time $t_2=t_1 + \delta t$ by
\begin{equation}\label{eq:deltaF}
\delta F = \delta v^a \{F, \phi_a\}
\end{equation}
where the $\phi_a$ range over the complete set of first-class primary constraints and the $v^a$ are the totally arbitrary part of the $u^m$, with $\delta v^a = (v^a-\tilde{v}^a) \delta t$ where $v^a$ and $\tilde{v}^a$ are two different choices of $v^a$ at $t_1$.\footnote{Cf.\ \citet[\S1.2.1]{hentei}.} In a deterministic theory, the transformation (\ref{eq:deltaF}) does not modify the physical state and is thus considered a gauge transformation. In this sense, the first-class primary constraints generate gauge transformations. The famous `Dirac conjecture' attempts to extend this result to include all first-class constraints as generating gauge. In general, however, the conjecture is false as the existence of some admittedly contrived counterexamples illustrates.\footnote{Cf.\ \citet[\S1.2.2]{hentei}.} There is no harm for present purposes, however, if we assume that all first-class constraints generate gauge transformations. The restriction of a phase space function $F$ to $\mathcal{C}$ is gauge-invariant just in case $\{F,\phi_a\}\approx 0$, in which case (\ref{eq:deltaF}) implies $\delta F \approx 0$. The first-class constraints are thus seen to generate motions within $\mathcal{C}$. In contrast, second-class constraints generate motions leading outside of $\mathcal{C}$.\footnote{Cf.\ \citet[\S10.2.2]{belear01}.} This distinction permits the explication of another important concept: the gauge orbit. A {\em gauge orbit} is a submanifold of $\mathcal{C}$ which contains all those points in $\mathcal{C}$ which form an equivalence class under a gauge transformation. The sets of these points are path-connected in $\mathcal{C}$ since gauge transformations that connect these points are continuous and do not leave $\mathcal{C}$. They form a curve in $\mathcal{C}$. The gauge motion produced by the first-class constraints can thus be seen to be the tangents to these curves. The points of the gauge orbits in $\mathcal{C}$, equipped with a projection $\mathcal{C}\rightarrow \Gamma_{phys}$, constitute the so-called {\em reduced} or {\em physical phase space} $\Gamma_{phys}$. The physical phase space $\Gamma_{phys}$ is defined as the set of points representing gauge equivalence classes of points in $\Gamma$. In other words, the physical phase space is obtained by identifying all points on the same gauge orbits. This means that the bundle of admissible dynamical trajectories passing through a particular point $x\in \mathcal{C}$ is mapped to the physical phase space such that the bundle is projected onto a single dynamical trajectory through the point in $\Gamma_{phys}$ representing the gauge equivalence class in which $x$ falls.

Assume a Hamiltonian system with constraints is given. Assume further that all constraints are first-class.\footnote{Second-class constraints can be regarded as resulting from fixing the gauge of a `larger' system with an additional gauge invariance. They can be replaced by a corresponding set of first-class constraints which capture the additional gauge invariance. Second-class constraints are thus eliminable. In fact, in some cases, it may prove advantageous to thus `enlarge' a system as this permits the circumvention of some technical obstacles \citep[\S1.4.3]{hentei}, albeit at the price of introducing new `unphysical' degrees of freedom. Without loss of generality, we can thus consider a Hamiltonian system whose constraints are all first-class.} Constraint equations are equations which the canonical variables must satisfy in addition to the dynamical equations of the system. If a set of variables were to determine one and only one physical state, then, given the existence and uniqueness of the solutions of the dynamical equations, one could plug the set of variables uniquely specifying the state into the dynamical equations and could thus obtain the full deterministic dynamical evolution of the physical degrees of freedom. If constraints are present, however, a set of variables does not uniquely describe a physical state. Solving the constraints thus means to use these additional equations to explicitly solve for a variable. This permits the elimination of this variable (and the now solved constraint equation). Solving the constraints of the constrained Hamiltonian system thus amounts to the reduction of the number of variables used to specify the physical state of the system. Once all constraint equations are solved and thus eliminated, the remaining canonical variables are ineliminable for the purpose of uniquely specifying a physical state. In this case, we are back to an unconstrained Hamiltonian system in the sense that its phase space is its {\em physical} phase space. In the absence of any second-class constraints, the total number of canonical variables $(=2N)$ minus twice the number of first-class constraints equals the number of {\em independent} canonical variables. Equally, the number of {\em physical} degrees of freedom is the same as half the number of independent canonical variables, or the same as half the number of canonical variables minus the number of first-class constraints.\footnote{This manner of counting the physical degrees of freedom is well defined for any finite number of degrees of freedom, and perhaps for countably many too. For uncountably many degrees of freedom, new subtleties arise. Cf.\ \citet[\S1.4.2]{hentei}.}

\subsection{Gauge freedom in Hamiltonian general relativity}\label{ssec:hamgr}

Hamilton's equations, at least in the narrower standard sense, explicitly solve for the time derivatives. This can only be achieved within GR if its original 4-dimensional quantities are broken up into (3+1)-dimensional quantities, with time accruing in the one single dimension. Similar coercion must be exercised upon the four-dimensional structure of spacetime, nota bene, when we wish to consider an initial-value formulation of GR. In order to find a Hamiltonian or an initial-value formulation, GR must be regarded as describing the dynamical evolution {\em of} something. Breaking up spacetime into `space' that evolves in `time' in order to determine whether a well-posed initial-value formulation exists, i.e.\ whether the physical degrees of freedom enjoy an at least minimally stable deterministic evolution, becomes manageable once we impose a gauge condition to weed out any unphysical degrees of freedom. The traditional formulation of GR as a constrained Hamiltonian system entertains twelve dynamical variables, the six independent components of the three-metric $q_{ab}$ and the six independent components of the corresponding conjugate momentum $\pi^{ab}$. Half this number is six, and there are four first-class constraint equations, which leaves the gravitational field with two physical degrees of freedom per point in space. Fortunately, this is the same number of degrees of freedom as one gets for a linear spin-2 field propagating on a flat spacetime background, which can be considered as a weak-field limit of GR.\footnote{See \citet[\S4.4b]{walgr}; cf.\ also \citet[266]{walgr} for a slightly different way of calculating the degrees of freedom of the gravitational field.} With a gauge condition enforced, Einstein's field equations can be massaged into a form of hyperbolic second-order differential equations defined on manifolds which admit existence and uniqueness theorems. Even in an appropriate gauge fix, however, GR allows for ways in which the field equations may fail to uniquely determine their solutions.\footnote{For an explanation of the failures of determinism in this setting, cf.\ \citet[\S4.1]{wut06}, on which the past few pages have been based. Also, and at the peril of burying an absolutely central point in a footnote, this severance of space and time threatens the general covariance so central to GR. How general covariance gets implemented in Hamiltonian GR and the subtleties that arise in doing so are discussed in \citet[\S4.4]{wut06}. What follows explicates the gist of this implementation.} 

The conceptually most momentous consequence of casting GR as a constrained Hamiltonian system is that the Hamiltonian $H$ is itself a constraint bound to vanish on the constraint surface of the phase space. This is what ultimately leads to the `problem of time', a conceptual tangle in the foundations of Hamiltonian GR and of quantizations relying thereon, consisting of essentially two strands, the disappearance of time as a fundamental magnitude and the `freezing' of the dynamics. The first aspect, the vanishing of time as a fundamental physical magnitude, is suggested at the classical level by the increasing elimination of time in classical physics, leading up to Hamiltonian GR, as it is retraced in \citet[\S\S2.1 and 2.2]{hugeal13}. However, there is a sense in which it only comes to full fruition in quantum theories, as will be elaborated below. 

The freezing of the dynamics---more aptly called the `problem of change'---, however, fully appears at the classical level. A crucial premise of the argument leading to the problem of change is that only gauge-invariant quantities can capture the genuinely physical content of a theory. This premise is justified by pointing to the fact that two distinct mathematical models of a theory describe the same physical situation just in case they are related by maps which are interpreted as `gauge' transformations. Of course, it may be controversial for any given theory just which maps ought to be considered `gauge', but I take the justificatory fact invoked in the previous sentence to be analytic of what it means to be `gauge', viz.\ to capture a representational redundancy not reflective of the true physical situation. In other words, the premise stipulates that the physical content of a theory is exhausted by the gauge-invariant quantities as codified by the theory. The concept of `Dirac observables' tries to capture this idea in the context of constrained Hamiltonian theories:
\begin{defi}[Dirac observables]\label{def:dirobs}
A(n equivalence class of) {\em Dirac observable(s)} is defined as the (set of those) function(s) in phase space that has (have) weakly vanishing Poisson brackets with all first-class constraints (and coincide on the constraint surface). Equivalently, Dirac observables are functions in phase space which are constant along gauge orbits on the constraint surface.
\end{defi}
Thus, if the premise is true, and if the gauge-invariant quantities of a constrained Hamiltonian theory are precisely its Dirac observables as defined in Definition \ref{def:dirobs}, then the physical content of a constrained Hamiltonian theory is exhausted by its Dirac observables. 

In order to determine the physical content of Hamiltonian GR, thus, it becomes paramount to identify its first-class constraints. I will not execute this task here with the mathematical precision it deserves but rest content with a conceptual motivation.\footnote{For a somewhat rigorous execution in the case of the so-called ADM and Ashtekar-Barbero versions of Hamiltonian GR, cf.\ \citet{wut06}, \S 4.2.1 and \S4.2.2, respectively.} The vantage point is the principle of general covariance so central to GR. This principle demands that the Einstein equations' dynamical symmetry group {\em Diff$(\mathcal{M})$} of active spacetime diffeomorphisms is the gauge group of GR.\footnote{A {\em spacetime diffeomorphism} is a one-to-one and onto $C^\infty$-map from $\mathcal{M}$ onto itself which has a $C^\infty$-inverse. Diffeomorphisms induce transformations in the fields defined on the manifolds. Intuitively, a map between manifolds is active if it `moves around' the points without recourse to any coordinate system. Thus, an active transformation is not a change in coordinate systems, but a transformation pushing around the physical fields on the manifold. But this metaphorical picture should be enjoyed with the adequate mathematical caution.}\footnote{This is the received view, but it should be noted that there has been recent dissent, e.g.\ in \citet[\S3]{cur09}.} In other words, active spacetime diffeomorphisms, which map a solution of the dynamical equation to another solution, ought to be considered relating two mathematically distinct solutions describing one and the same physical situation.\footnote{For a detailed analysis and justification, cf.\ \citet[\S3, particularly \S3.2]{wut06}.} Thus, general covariance is spelled out as gauge invariance under active spacetime diffeomorphisms. 

In the Hamiltonian formalism, the dynamical symmetry of GR gets encoded as constraints which generate the spacetime diffeomorphisms in the sense explained in \S\ref{ssec:hamcon}. In the standard formulation of GR, the elements of the symmetry group {\em Diff$(\mathcal{M})$} are defined as maps between four-dimensional manifolds. The Hamiltonian formalism breaks this four-dimensionality down to a three-plus-one-dimensional rendering; accordingly, {\em Diff$(\mathcal{M})$} breaks down into a group of three-dimensional `spatial' diffeomorphisms and a group of one-dimensional `temporal' diffeomorphisms. This move is not without subtleties, as expounded in \citet[\S4.2]{wut06}: the symmetry group in Hamiltonian versions of GR differs from that in the usual articulation of the theory, thus distinguishing Hamiltonian GR from its standard cousin in yet another way from those given at the end of the section. In the exemplary ADM version of Hamiltonian GR, the spacetime diffeomorphisms are generated by normal and tangential components of the Hamiltonian flow. Since the constraints generating the diffeomorphism must vanish (weakly), these components of the Hamiltonian vanish (weakly). Furthermore, in a Hamiltonian theory, it is the Hamiltonian which generates the dynamical evolution via the Hamilton equations. Since the Hamiltonian is constrained to vanish, the dynamics gets `frozen'. 

More specifically, (the normal component of the) Hamiltonian is a first-class constraint. Thus, the Dirac observables must have weakly vanishing Poisson brackets with the Hamiltonian and thus turn out to be constants along the gauge orbits generated by the Hamiltonian. This accords with the stipulation above that the physical-content-capturing Dirac observables must be invariant under gauge transformations, here constituted by active spacetime diffeomorphisms. Since the Dirac observables are constant along orbits generated by the Hamiltonian, all genuinely physical magnitudes must be constants of the motion, i.e., they must remain constant over time. In other words, any supposed change is purely a representational redundancy, and not a physical fact. Thus, the argument concludes, there is no change! Since GR, or any quantum theory of gravity replacing it, is a fundamental theory, we are saddled with the uncomfortable task of explicating how time and change can arise phenomenologically---which they undoubtedly do---in a fundamentally changeless world. {\em O quam cito transit gloria temporis}.\footnote{For a discussion of philosophical reactions to this situation, cf.\ \citet[\S2.3]{hugeal13}.}

Avoiding this unpalatable conclusion might be all too easy by simply brushing aside Hamiltonian GR as a failed articulation of the theory. But this move is not readily available, at least not without some considerable cost. A prima facie justification for brushing it aside points out that Hamiltonian GR is not theoretically equivalent to the standard formulation of GR. It is true: Hamiltonian GR presupposes that spacetimes can always be sliced up to conform to its $(3+1)$-dimensional framework, but this is demonstrably false in GR. Thus, Hamiltonian GR at best captures the sector of GR containing sliceable, globally hyperbolic spacetimes. Furthermore, known articulations of Hamiltonian versions of GR exclude any matter content from the spacetimes and thus only codify vacuum spacetimes. It is not clear, however, that this inequivalence suffices to evade the strictures of the above argument. And most importantly, Hamiltonian formulations of GR serve as the asis for one of the most important family of approaches to formulating a quantum theory of gravity. By virtue of this fact alone, they deserve to be taken seriously, not just mathematically, but also philosophically.

\section{How spacetime dissolves in LQG}\label{sec:dissolve}

Once the classical theory is cast in a Hamiltonian fashion, then it can be subjected to the powerful canonical quantization technique. This procedure, pioneered by Paul Dirac, converts the canonical variables of the classical theory into quantum operators defined on an appropriately chosen Hilbert space. The Poisson bracket structure of the classical level is thereby transposed to give rise to the canonical commutation relations obtaining between the basic operators in the quantum theory. From these basic operators, more complex operators can be built up. The classical constraint functions get translated into such complex operators acting on elements in the Hilbert space, thus turning the constraint equations into wave equations. Since they are constraint equations, the constraint operators annihilate the states on which they are acting. Only those states which are so annihilated by the constraints operators are considered {\em physical} states. As usual in quantum mechanics, the Hamiltonian operator $\hat{H}$ generates the dynamics via a Schr\"odinger-type equation. 

As we have seen in \S\ref{ssec:hamgr}, in Hamiltonian formulations of GR, the Hamiltonian itself becomes a constraint. In the quantum theory, we get
\begin{equation}\label{eq:wdw}
\hat{H} |\psi\rangle = 0
\end{equation}
which is demanded to hold for all physical states $|\psi\rangle$. The `physical' Hilbert space $\mathcal{H}$ consists just of those states, which satisfy all constraints, i.e., are annihilated by all constraint operators in the theory. Equation \eqref{eq:wdw}, also called the `Wheeler-DeWitt equation', gives a very direct intuition of both the problem of time and that of change. Concerning the problem of time strictly so called, comparing \eqref{eq:wdw} to the ordinary Schr\"odinger equation,
\begin{equation}\label{eq:schroed}
\hat{H}|\psi\rangle = i \hbar \frac{\partial}{\partial t} |\psi\rangle,
\end{equation}
we notice the absence of the time parameter $t$ in \eqref{eq:wdw}. This is indicative of the problem of time: the absence of time from the fundamental picture. Quite literally, time drops out of the equation in Hamiltonian quantum gravity.

Given that \eqref{eq:wdw} plays the role of the dynamical equation in quantum Hamiltonian GR just as \eqref{eq:schroed} does for ordinary quantum mechanics, we also glean the first traces of the quantum version of the problem of change by recognizing that the time derivative vanishes. Analogous to the classical case, constraint operators generate the gauge symmetries of the theory. Accordingly, the criterion for the gauge-invariant observables, the Dirac observables defined in Definition \ref{def:dirobs} of the quantum theory, gets translated as requiring that functions $\hat{F}$ of operators represent Dirac observables just in case they commute with all the constraint operators $\hat{C}_i$
\begin{equation*}
[\hat{F}, \hat{C}_i] |\psi\rangle = 0,
\end{equation*}
for all $i=1,..., m$, where $m$ is the number of constraints, and for all $|\psi\rangle$ in $\mathcal{H}$. This entails that every Dirac observable must commute with the Hamiltonian. Since the Hamiltonian is what generates the dynamical evolution of the states, all Dirac observables must thus be constants of the motion, i.e., not changing over time. However, the Dirac observables also exhaustively capture the physical content of the theory, at least according to the premise stated in \S\ref{ssec:hamgr}. Thus, no genuine physical magnitude changes over time. Hence, the dynamics of the world described in canonical quantum gravity is `frozen' in time. There simply is no change at the most fundamental level described by these Hamiltonian quantum theories of gravity! Change, as it turns out, only arises as a representational artefact---`gauge'---with no physical counterpart in the fundamental theory. 

Unlike at the classical level, where arguably the strictures of the argument can be evaded, at least to some extent, by avoiding Hamiltonian formulations of GR, this is evidently not possible for quantizations based on them as the problem is built right into the framework. Perhaps we ought to have expected such an outcome---after all, GR teaches us that time is not external to the physical systems of interest but itself partakes as part of spacetime in dynamical interactions with the material content of the universe, which constitute the usual physical systems physics describes. In other words, time is part of the physical system we are trying to quantize. 

In fact, indications persist that quite generically in quantum gravity space and time, at least as standardly understood in GR, no longer form part of the fundamental ontology. Instead, space and time, or at least one or the other, are `emergent' phenomena that arise from the basic physics. As it is used in the present essay, `emergent' should not be taken as the terminus technicus in philosophy that designates properties which are not even weakly reducible. Rather, it should be considered as an umbrella term for a relationship that may well turn out to be reductive, as will be argued in \S\ref{ssec:nonred}. In fact, to characterize the exact nature of this relationship is the ultimate goal of the research addressing the issue at stake. In the language of physicists, spacetime theories such as GR are `effective' theories trading in `emergent' phenomena, much like thermodynamics is an effective theory dealing with the emergent phenomenon of temperature, as it is built up from the collective behaviour of gas molecules. However, quite unlike the fact that temperature is emergent, the idea that the universe and its material content is not {\em in} space and time shocks our very idea of physical existence as profoundly as any previous scientific revolution did.

So there is at least a sense in which time vanishes in canonical approaches to quantum gravity. It has been argued that because string theory contains GR ``in some limit... [t]he disappearance of external time should... also hold in string theory'' \citep[10]{kie12}. As a consequence of the holographic principle, space as well can be considered emergent in string theory \citep{hugeal13}. Furthermore, the fundamental structures postulated by various quantum theories of gravity diverge significantly from the familiar spacetimes of GR. For instance, so-called non-commutative geometry replaces the basic geometric picture we have of spacetime by algebraic relations between temporal and spatial coordinates or directions and generalizes multiplicative relations among them so that they no longer commute. This generalization has weird consequences and renders the basic structure conceptually quite different from spacetime \citep{hugeal13}. As another example, the fundamental structure generically turns out to be discrete rather than continuous. For a vast class of quantum theories of gravity, Lee Smolin, takes discreteness to be ``well established.'' \citeyearpar[549]{smo09}. 

Of course, one might react to these developments as John Earman did, at least concerning LQG, and insist that
\begin{quote}
although classical general relativistic spacetime has been demoted from a fundamental to an emergent entity, spacetime per se has not been banished as a fundamental entity. After all, what LQG offers is a quantization of classical general relativistic spacetime, and it seems not unfair to say that what it describes is quantum spacetime. This entity retains a fundamental status in LQG since there is no attempt to reduce it to something more fundamental. \citeyearpar[21]{ear06a}
\end{quote}
If this is just a quarrel over words, I have no appetite to engage in it. We are free to call LQG's fundamental structure, to be described in the remainder of this section, `quantum spacetime' all right, but given the profound departures from relativistic spacetimes, the use of a different term is not only warranted, but also preferable, as I have argued elsewhere \citep{wut12b}. Let us leave this debate to one side and delve into the physics in order to get a sense of what it is LQG theorizes about.

\subsection{Introducing LQG}\label{ssec:lqg}

Canonical quantum gravity generally, and LQG in particular, attempt to transpose the central lesson of GR into a quantum theory. The pertinent key innovation of GR is the recognition that spacetime does not passively offer a fixed `background' which determines the inertial `forces' acting on the physical content of the universe, but instead a dynamical structure which interacts with matter. To repeat, LQG is based on a reformulation of GR as a `Hamiltonian system', which reinterprets spacetimes as $(3+1)$-dimensional rather than $4$-dimensional, with constraints. Thus, recasting GR as a Hamiltonian theory forces a `foliation' of its spacetimes by an equivalence relation into three-dimensional `spatial' hypersurfaces, parametrized by a one-dimensional `time'. The natural interpretation of the Hamiltonian system would be that of a three-dimensional `space' considered as a dynamical physical system which evolves over `time', where the three-dimensional hypersurfaces would represent the instantaneous state of the dynamical theory.

LQG is thus a canonical quantization of Hamiltonian GR.\footnote{For a thorough introduction to LQG, cf.\ \citet{rov04}; for the mathematical foundations, cf.\ \citet{thi07}. \citealt{rov11a} is a recent review article.} Before we proceed, let it be noted that the particular formulation required entails a substantive limitation of the approach: only `globally hyperbolic' spacetimes of the classical theory are considered. If a spacetime is globally hyperbolic, then it is topologically `$3+1$', i.e., the topology of $\mathcal{M}$ is $\Sigma\times\mathbb{R}$, where $\Sigma$ is a three-dimensional submanifold of $\mathcal{M}$.\footnote{For a more systematic explication of global hyperbolicity and neighbouring concepts, see \citet[593]{smewut11}.} Just how severe this limitation is is debatable; many physicists do not consider it troubling, some philosophers have dissented. To impose global hyperbolicity as a necessary condition for physically reasonable spacetimes amounts to asserting a strong form of the merely conjectured, but not proven, cosmic censorship hypothesis. Dissenting voices cautioning against stipulating global hyperbolicity as necessary include \citet{ear95}, Erik \citet{cur01}, Chris \citet{smewut11}, and John \citet{man11b}. \citet[414]{man11b} proves that as long as a spacetime is not ``causally bizarre'', it is observationally indistinguishable from another spacetime, not isometric to the first and not globally hyperbolic, yet with exactly the same local properties. From this, Manchak concludes that ``[i]t seems that, although our universe may be... globally hyperbolic..., we can never know that it is.'' (ibid.) In the light of this result, it appears brash to enthrone global hyperbolicity as a sine qua non of physical reasonability. Having said that, however, if LQG were to be a huge empirical success, its premises would be vindicated. Note the future subjunctive tense in the previous sentence.

There currently still persists another, uncontroversially problematic, limitation of the approach: only vacuum spacetimes are considered, i.e., the classical vantage point of the approach is the vacuum sector of GR with everywhere vanishing energy-momentum. This technical simplification comes at the price of rendering it unclear whether the resulting quantum theory can deal with a non-zero energy and matter content of the universe, presumably a necessary condition for giving an empirically adequate account of the actual world. The situation may not be quite as bleak for LQG as this may suggest, for three reasons. First, vacua are physically important states and their theoretical understanding may shed decisive light on the necessary steps leading to a more general theory encompassing matter. Secondly, the assessment as to whether or not models of a theory or vacuum states of the universe contain matter may come apart for classical and quantum theories. In other words, the quantum theory which started out from classical vacuum states may be interpreted to contain matter. This possibility does not come without further complications, though: the emerging matter may well be highly non-local and may violate most or all energy conditions. Thirdly, and most speculatively, matter, just as space and time, may emerge from the---perhaps topological or combinatorial---properties of the fundamental structure and hence not be present at the fundamental level. 

The goal of the quantization is to find the Hilbert space corresponding to the physical state space of the theory and to define operators on the Hilbert space representing the relevant physical magnitudes. The hope would naturally be that some of the latter make contact to the empirically testable. In order to get the quantization started, one chooses a pair of canonically conjugate variables which coordinatizes the relevant sector of the classical phase space. Different choices lead to different quantum theories: geometrodynamics's choice is the induced three-metric on the three-hypersurface and its conjugate momentum constructed from the external curvature of the three-hypersurface, LQG starts out from Abhay Ashtekar's `new variables' of a connection $A^i_a$ and its conjugate, a densitized triad `electric field' $E_i^a$ and constructs a `holonomy' and its conjugate `flux' variables from them. The geometrical structure of the classical phase space is encapsulated in the canonical algebra given by the Poisson brackets among the basic variables. This structure gets transposed into a quantum theory by first defining an initial functional Hilbert space of quantum states $|\psi\rangle$. The basic canonical variables are turned into operators whose algebra is determined by their commutation relations arising from the classical Poisson brackets. The classical constraints, which are functions of the canonical variables, now become operators constructed `isomorphically' as functions of the basic operators. Classically, the constraint functions are set to zero; in the quantum theory, they annihilate the states. Thus, by imposing the constraints, the theory effectively demands that only states which are annihilated by {\em all} constraint operators are considered physical. Dynamical equations, as was already clear at the outset of this section, play a somewhat different role. In a sense, given that the `Schr\"odinger-like' equation of the quantum theory is the constraint equation \eqref{eq:wdw}, there is no additional dynamical equation governing any `dynamics' of the theory.

In LQG, three families of constraints arise. First, the so-called `Gauss constraints' indicate a rotational gauge freedom of the triads and generate an infinitesimal $SU(2)$ transformation in the internal, as opposed to spacetime, indices (indicated by letters from the middle of the alphabet). These are comparatively straightforward to solve. Next, we find three `(spatial) diffeomorphism' constraints, which generate the spatial diffeomorphisms on the three-hypersurfaces. These constraints are hard to solve, but it has been done. The resulting Hilbert space, i.e., the Hilbert space we obtain from the states which get annihilated by the Gauss and diffeomorphism constraints, is called the `kinematical Hilbert space' and will here be denoted by $\mathcal{H}_K$. Finally, there is the Hamiltonian constraint which has so far defied solution. In fact, it is not even clear what the concrete form of the formal equation \eqref{eq:wdw} is. In this sense, LQG is not yet a complete theory. As will hopefully become clear later in the essay, there remain plenty of reasons not to walk away from LQG, at least not just yet.

Given the technical and conceptual difficulties with the `dynamics' \eqref{eq:wdw}, various authors have sought ways to circumvent the standard conceptualization of dynamics in a Hamiltonian theory. One main approach conceives of the dynamics in ways similar to perturbative approaches to quantum field theory, taking elements of $\mathcal{H}_K$ as three-dimensional `initial' and `final' `spaces' and compute transition amplitudes between them \citep[\S 3]{rov11a}. Or alternatively, as Carlo Rovelli has suggested, the states in the physical Hilbert space may not be `states at some time'; instead, they are `boundary states', i.e., states describing quantum space surrounding a four-dimensional region of spacetime.\footnote{A more detailed analysis of dynamics in LQG can be found in \citet[\S5.3]{wut06}.} 

Because \eqref{eq:wdw} is not solved yet, all results must remain preliminary. One way to see this immediately is to remind the reader that all Dirac observables must commute with all constraints. If we accept that the set of Dirac observables is identical to the set of genuine physical magnitudes, as arguably we should on pain of introducing gauge-dependent quantities, then we cannot determine the physical magnitudes yet, as we don't know the explicit form of $\hat{H}$ and so cannot determine which operators commute with it. It thus remains open whether any of the geometric operators to be introduced shortly really corresponds to a genuine physical magnitude. 

\begin{figure}
\centering
\epsfig{figure=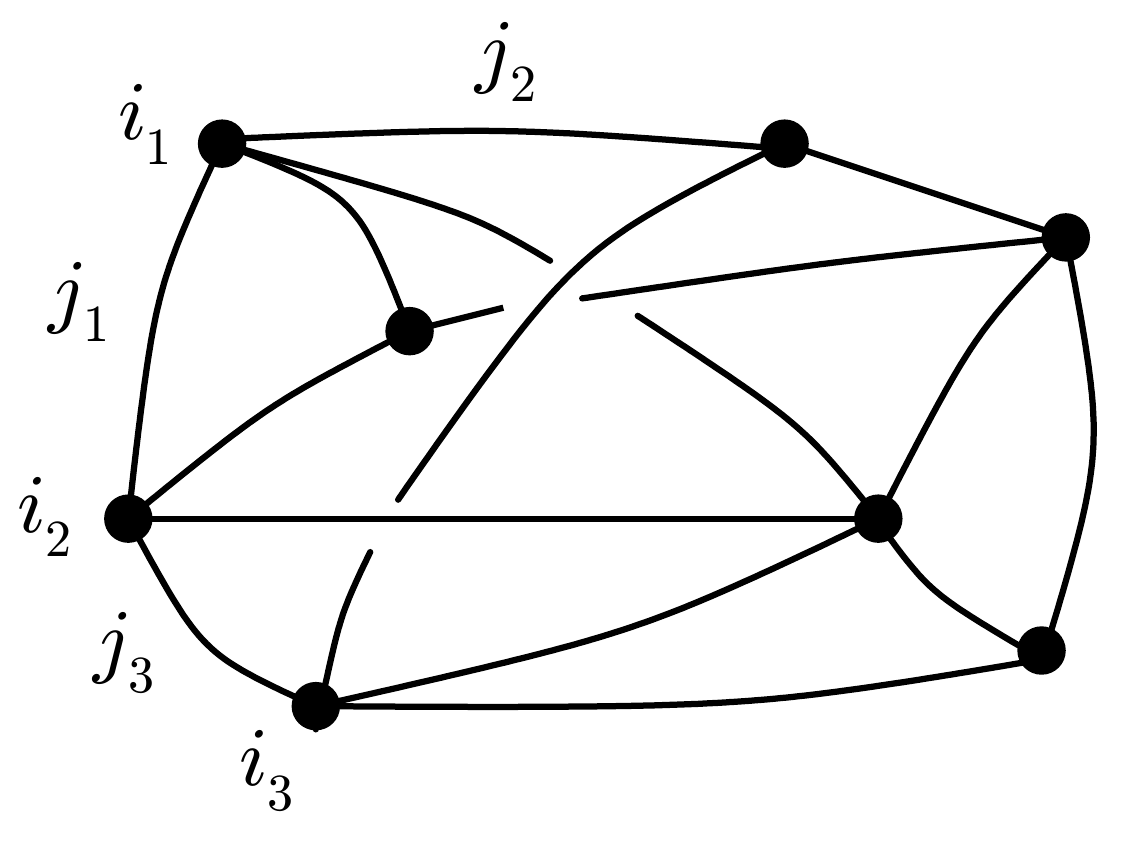,scale=0.55}
\caption{\label{fig:spinnetwork}A spin network state is characterized by an abstract graph with `spin'-representations on the nodes and the links between them.}
\end{figure}

Let us study the structure of $\mathcal{H}_K$ then. It turns out that so-called `spin network states' provide a useful basis in $\mathcal{H}_K$.\footnote{For the technical background of this basis and its interpretation, cf.\ \citet[\S2.3]{rov11a}} These spin network states are interpreted to be the quantum states of the gravitational field. Since physical `space' will be in a state in $\mathcal{H}_K$, as \S\ref{ssec:applying} will suggest, it will generally be in a quantum superposition of spin network states. Spin network states can be represented by {\em abstract labelled graphs} as in Figure \ref{fig:spinnetwork},\footnote{More precisely, they are represented by labelled graphs embedded in some background space. Thus, they are not invariant under spatial diffeomorphisms, i.e., when they are `pushed around' on the embedding manifold. In order to fully solve the diffeomorphism constraints, then, we need {\em equivalence classes} of spin network states under three-dimensional diffeomorphisms on the background manifold. Sometimes, these equivalence classes, represented by {\em abstract labelled graphs}, are called `$s$-knot states' in the literature. So I am being slightly sloppy by using the locution `spin network states' ambiguously.} as they are completely characterized and uniquely identified by three types of `quantum numbers'. 
The first label characterizes the abstract graph $\Gamma$, the second the irreducible $SU(2)$-(hence `spin') representations $j_l$ on the links and a third the $SU(2)$-representations on the nodes, denoted by $i_n$. It should be emphasized that the abstractness of the graph is central to the correct interpretation of the emerging picture here: the spin network states are not quantum states of a physical system {\em in} space; rather they are the quantum states {\em of} physical space.

The spin network states $|\Gamma, j_l, i_n\rangle$ are eigenstates of the so-called area and volume operators defined on $\mathcal{H}_K$. The spectra of these operators yield important information concerning the geometrical interpretation of the spin network states, although it must be emphasized that the interpretation of the states relies, in turn, on an {\em interpretation} of these operators as geometric. Since we study the properties of the gravitational field via the geometry of the physical space, the properties of (three-dimensional) gravitational fields are determined by the spectral properties of the area and volume operators. These operators, which will be discussed in greater detail in \S\ref{ssec:applying}, turn out to have discrete spectra \citep{ashlew97,ashlew98,ashlew99,rovsmo95a,rovsmo95b}. The granularity of the spatial geometry---the `polymer' geometry of space---follows from the discreteness of the spectra of the volume and the area operators. Essentially, each node (and only the nodes) in the network contributes a term to the sum of the volume of a region. On each node, there sits an `atom' of space with volume $\mathbf{V}_n$, as it were. These elementary grains of space are separated from each other by their surfaces of contiguity. Just as the volume operator receives contributions from the nodes of a region, the area operator acquires contributions from all the links that intersect the surface. For instance, the surface whose only intersecting link is a link with quantum number $j_l$ has a surface area of $\mathbf{A}_l\propto\sqrt{j_l(j_l + 1)}$ \citep[\S6.7]{rov04}. Thus, the `size' of the surface connecting adjacent `chunks' of `space' is constructed from the spin representations sitting on the relevant links. Thus, the smooth space of the classical theory is supplanted by a {\em discrete quantum structure} displaying the granular nature of space at the Planck scale. Continuous space as we find it in classical theories such as GR and as it figures in our conceptions of the world is a merely {\em emergent phenomenon}.\footnote{It should be kept in mind, however, that these operators are not Dirac observables and should therefore be taken with a grain of salt. They are partial observables in the sense of \citet{rov02}.}

Physical three-space, in Rovelli's interpretation, is a quantum superposition of spin network states, analogously to the physical electromagnetic field consisting of a superposition of $n$-photon states. LQG {\em predicts} the existence of indivisible quanta of volume, area, and length, as well as their spectra (up to a constant). Importantly, this discreteness was a {\em result} of the loop quantization, rather than an {\em assumption}. According to LQG, measurements of the Planck geometry of space must therefore yield one of the values in the spectrum of the concerned operator.

As mentioned above, the `dynamics' of canonical LQG are only known in formal outline. As in any Hamiltonian theory, the dynamics of the theory is generated by the Hamiltonian operators $\hat{H}$, which is defined on $\mathcal{H}_K$, via the Wheeler-DeWitt equation \eqref{eq:wdw}. The space of the solutions of \eqref{eq:wdw} will constitute the {\em physical Hilbert space} $\mathcal{H}$. But since there exist several inequivalent versions of $\hat{H}$---all of which may be false---the Hilbert space $\mathcal{H}$ has not yet been constructed and the theory remains incomplete. 

Before we start to consider how spacetime emerges from the fundamental structures of LQG---spin network states---, let us make sure that it has indeed vanished from the fundamental ontology. Of course, as Earman suggested in the quote above, we might simply call the spin-network structure `quantum spacetime' and move on with it. To use homonyms, or near-homonyms, for two rather different structures, however, promises to create more confusion than comprehension. The spin networks diverge from classical relativistic spacetimes in at least two crucial points. First, unlike the continua of classical physics, they are discrete. As was observed above, many expect the fundamental structure in quantum gravity to be discrete and this expectation is certainly borne out in many of the extant approaches. This is a significant departure, but may not sway everyone to discontinue considering the fundamental structure a `spacetime'.

Arguably, however, the deeper divergence from classical spacetimes arises from the `non-localities' that we find in spin networks (and in many other quantum structures).\footnote{Cf.\ e.g.\ \citet{marsmo07}.} How these fundamental structures can be `non-local' needs a bit of explaining, given that (non-)locality is a spatiotemporal, or anyway a spatial, concept. To appreciate the sense in which the spin networks do contain `non-localities', consider a fundamental relational structure consisting of a set of basal atoms, which exemplify, in pairs, a basal `adjacency' relation. Together with an intrinsic `valence' attributed to each of the atoms and each of the exemplified relations, this yields a connected structural complex of the kind we find in LQG. Contrast this with the spatiotemporal structure we find in GR, where the spatiotemporal, indeed metric, relations obtaining between the spacetime events give rise to a locality and neighbourhood edifice. Now, these two structures are supposed to be related by an emergence relation. More specifically, the idea is that the exemplified fundamental structure is related, in some limit or in some approximation or at some scale, to a relativistic spacetime. Given two particular structures related in this way, one can map the atoms of the fundamental structure onto events in the spacetime. What it then means to say that there are `non-localities' present in the fundamental structure is that some pairs of adjacent basal atoms, i.e., pairs of atoms exemplifying the fundamental adjacency relation, get mapped onto events in the spacetime which can be at arbitrarily large distances now as measured in the metric of the emerging spacetime.\footnote{Cf.\ Figure 1 in \citet{hugwut13b}.} Locality is notoriously tricky in GR, of course, but in globally hyperbolic relativistic spacetimes, a precise notion of locality is readily available. Given a possibly physically privileged foliation, a spatial metric is induced on the leaf containing the events, which are thus spatially related. This now permits an explication of locality e.g.\ in terms of convex spatial neighbourhoods of events. Thus, what is adjacent in the fundamental structure in general is not local or nearby in the emerging spacetime as judged by the latter's induced spatial metric. 

From the perspective of the emerging spacetime, the spin networks generally get the locality structure wrong, or so one would expect. The expectation that these non-localities are generic arises from the fact that relation between spin networks and classical spacetimes---to the extent to which we understand it---is many-to-one.\footnote{Cf.~Section \ref{sec:reemerge}. Cf.\ also \citet[\S2]{marsmo07} who give a related reason.} In other words, there are in general many spin network states whose best classical approximation is the same relativistic spacetime. Since these spin networks are physically distinct, and one of the main ways in which they can differ is by their connectivity defined by the obtaining adjacency relations, spin networks with distinct topologies will be best approximated by one and the same spacetime. As spin networks that give rise to realistically large universes will consist of very many adjacent pairs of nodes, it seem natural to think that at least some of them will be non-local in the present sense. If this is right, then non-localities generically arise in spin networks, and we have a second deep departure of the latter from relativistic spacetimes. 

These non-localities are suppressed in the low-energy approximation from the spin network to the relativistic spacetime. In fact, they must be suppressed, for otherwise they would have to be emulated by the emerging structure in the sense that these adjacency relations would re-occur in the spacetime in the form of neighbouring relations and thus not qualify as `non-localities'. To repeat, `non-localities' of the relevant sort are fundamental adjacencies with no vicinity-type counterpart in the emerging spacetime. If the course graining attendant to the emergence of spacetime from spin network states---of which more in \S\ref{sec:reemerge}---would not `wash out' the non-local connections, they would have to be encoded in the emerging relativistic spacetime, perhaps as non-local `wormholes'. If, however, their presence were so strong as to preclude essentially local physics at comparatively low energy scales, such as described by quantum field theory on relativistic spacetime backgrounds, then the corresponding theory, or at least model, would have to be considered empirically inadequate.\footnote{This does not entail that the fundamental non-localities could not have observable consequences, such as those proposed by \citet{presmo09}.} So we would expect those non-localities to be generically present, but suppressed in the coarse graining to macroscopic scales. 

Relativistic spacetimes arguably differ in significant ways in how they conceptualize space and time from our intuitive concepts of space and time. But whatever differences these are, they do not suffice to call into question why we refer to the structures of GR as `spacetimes', and justifiably so. Whatever the differences between intuitive space and time and spacetime in GR may be, it is clear that the departures of LQG from the manifest image run much deeper. Not only is the fundamental structure discrete and non-local, but as we have seen in \S\ref{sec:probtime}, the problem of time in its different forms illustrated how our common concepts of time, change, and dynamics and the way these concepts are standardly encoded in physical theories and their languages completely and utterly fails. Even though this failure was enunciated in \S\ref{sec:probtime} at the classical level already, it crucially depended on the particular non-standard, and inequivalent, formulation of GR necessary for the canonical programme to get going. If we could directly quantize GR from its standard formulation, the resulting theory's departure from classical spacetime physics might be milder. But alas, no promising strategy along these lines is known. 

I conclude that we can safely assume that spacetime has been lost, at least in its traditional, relativistic sense, somewhere in the transition from GR to LQG. Now that the Babylonians of quantum gravity have removed spacetime from its sacred place, amid rampant speculation concerning its whereabouts, serious efforts have commenced to recover the lost spacetime and restore it to its lawful place. He or she who recaptures it may be blessed with wisdom---or be smitten, as the case may be.

\section{What emergence of spacetime is not}\label{sec:nonem}

In order to honour the covenant---and to avoid being smitten---, then, let this section clarify what the emergence of spacetime could not be. First, \S\ref{ssec:nonred} explains the difference between the standard concepts of `emergence' as they figure in philosophy and physics, respectively, and states that it is the physicists' use that will be relevant for our purposes. Secondly, it will be argued in \S\ref{ssec:unitary} that the use of the notion of `unitary equivalence' will not serve to determine whether spacetime still maintains fundamental existence in LQG.

\subsection{Non-reductive relation}\label{ssec:nonred}

The concept of `emergence' has a venerable history in philosophy: arguably stretching back to Aristotle and Galen, it attracted renewed interest in the nineteenth century, reflected in the work of George Henry Lewes, John Stuart Mill, and C D Broad in Britain, and Nicolai Hartmann on the continent. Despite some variation among them, authors in this tradition as well as contemporary philosophers use the term so as to imply a {\em non-reductive relation} between the emergent and the fundamental, presupposing that reality is somehow layered into different `strata' and that the properties and relations attributed to entities at different levels in general differ from one another. The general spirit of the concept is well captured by Brian McLaughlin's definition in terms of supervenience:
\begin{defi}[Emergent property]\label{def:emergent}
``If P is a property of w, then P is emergent if and only if (1) P supervenes with nomological necessity, but not with logical necessity, on properties the parts of w have taken separately or in other combinations; and (2) some of the supervenience principles linking properties of the parts of w with w's having P are fundamental laws.'' \citep[39]{mcl97}
\end{defi}
Definition \ref{def:emergent} only gets traction if all the terms in the definiens are defined in their turn. Let us briefly discuss some of them. The first clause in the definition betrays the physicalist underpinnings of the version of emergentism which I assume here as standard. As \citet[\S3]{mcl97} explains, the relevant notion of `supervenience' in this context is based on the idea of a ``required-sufficiency relationship'' (ibid.), i.e., that the possessing of a higher-level property requires the possessing of a lower-level property which in turn suffices for the possessing of the higher-level property. This supervenience should not be forced by logic alone, but instead result from contingent laws of nature. To grasp the meaning and the role of the second clause, let me state the definition of `fundamental law' as given by McLaughlin:
\begin{defi}[Fundamental law]
``A law L is a fundamental law if and only if it is not metaphysically necessitated by any other laws, even together with initial conditions.'' (ibid., 39)
\end{defi}
The second clause is necessary; for without it, Definition \ref{def:emergent} would be overly inclusive, as McLaughlin argues, in that reducible properties would often also qualify as emergent, against the stated intention of the emergentists. If the laws which codify the connections between the properties of the lower-level entities with those of the higher-level, or those of the parts with those of the whole, are fundamental, then they are in principle not reducible to other laws governing the properties of lower levels, thus ruling out that reducible properties qualify.\footnote{For an up-to-date review on emergent properties, cf.\ \citet{ocowon12}.} 

It should be emphasized that in the context of the present study, and of much of the physics literature on the subject, `emergent' should not be understood as the terminus technicus defined in Definition \ref{def:emergent}, where an emergent property (or, mutatis mutandis, an emergent entity) is not even weakly reducible. Rather, it is to be understood as a collective designation for broadly reductive relationships. Indeed, {\em that} is the point of the entire enterprise: to understand how classical spacetime and its properties reduce, or more neutrally {\em relate}, to the fundamental non-spatiotemporal structure. Reduction, as an inter-theoretic relation, can thus be regarded as a {\em working hypothesis} of the quest to regain spacetime.

\subsection{Unitary equivalence}\label{ssec:unitary}

Leaving behind the general philosophical literature, we find in the pertinent philosophy of physics a very specific criterion which has been proposed to determine whether or not in a quantum theory of gravity spacetime can still be regarded as fundamental or not. Almost as an aside, Craig Callender and Nick Huggett \citeyearpar[21]{calhug01b} use the criterion of {\em unitary equivalence} for exactly this purpose, and in the context of LQG! Unitary equivalence, here as elsewhere, is used as a sufficient condition for physical equivalence. Callender and Huggett state that if bases of spin network states and of (functionals of) three-metrics in quantum geometrodynamics are unitarily equivalent, then they would merely constitute different representations of the same objects---viz.\ space---, rather than of numerically distinct objects. Hence, if successful, unitary equivalence would establish a particularly direct (reductive) relation, at least concerning {\em space}. If the two bases turn out to be unitarily inequivalent, then the reductive relation will be more complex. To invoke unitary equivalence as a (necessary and sufficient) condition for physical equivalence is well motivated.\footnote{At least at the level of ordinary quantum mechanics; in relativistic quantum theories, matters become more subtle. Cf.\ \citet[\S2.2]{rue11}.} Despite qualms one might entertain regarding the equivalence of the equivalences, let us grant, for the sake of argument, that unitary equivalence and physical equivalence come together. It turns out, however, that the criterion is nevertheless unhelpful, for three reasons. 

Since unitary (in)equivalence is usually predicated of {\em representations}, not of {\em bases}, let us translate the condition into the language of bases of Hilbert spaces before we start listing the problems:
\begin{defi}[Unitary equivalence between bases]\label{def:uniteq}
Two bases $\{|a^{(k)}\rangle\}$ and $\{|b^{(l)}\rangle\}$ of two Hilbert spaces $\mathcal{H}$ and $\mathcal{H}'$, respectively, are {\em unitarily equivalent} just in case there is a unitary map $U: \mathcal{H}\rightarrow \mathcal{H}'$ such that \ $U|a^{(k)}\rangle = |b^{(k)}\rangle$ for all $k$.
\end{defi}
Now, given this definition, and the orthonormality and the completeness of bases, it is easy to construct such a unitary map between Hilbert spaces of the same dimension:  $U = \sum_k |b^{(k)}\rangle\langle a^{(k)}|$. For our discussion below, we need to put two theorems on the table. Here is the first one:

\begin{theorem}[\citealt{debmik99}, 3.11.3(a)]
If $\mathcal{H}$ is an infinite-dimensional separable Hilbert space, then it is isomorphic to $l^2$, the space of square-summable sequences. 
\end{theorem}
Two Hilbert spaces are {\em isomorphic} just in case there is a unitary map that leaves the inner product invariant. Since being isomorphic is a transitive relation, {\em any} two infinite-dimensional separable Hilbert spaces are isomorphic. In other words, there is a unitary map between the bases of any two infinite-dimensional separable Hilbert spaces. This entails, of course, that for any two infinite-dimensional separable Hilbert spaces, we can find unitarily equivalent bases in the sense of Definition \ref{def:uniteq}. In fact, we have the more general theorem:

\begin{theorem}[\citealt{hal51}, \S16]
Any two Hilbert spaces $\mathcal{H}$ and $\mathcal{H}'$ are isomorphic iff $\dim(\mathcal{H}) = \dim(\mathcal{H}')$.
\end{theorem}
An immediate consequence of this theorem is that any two Hilbert spaces of the same dimension will have unitarily equivalent bases. So our knee-jerk reaction right after Definition \ref{def:uniteq} stands vindicated. Quite generally, the theorem shows that Hilbert spaces of the same dimension are geometrically indistinguishable and can thus rightfully be considered identical as far as their physically salient structure is concerned. 

Let us return to the proposal by \citet{calhug01b} and discuss its problems. As announced above, there are three of them. {\em Primo}, in order for this criterion to get any traction, the relevant Hilbert spaces would have to be known---but they are not. We have already seen that the physical Hilbert space $\mathcal{H}$ of LQG has not yet been constructed, only its kinematic Hilbert space $\mathcal{H}_K$. The same is true for geometrodynamics, where the constraints are non-polynomial and so far defy solution. No Hilbert space, no basis. No basis, no checking for unitary equivalence. But let us proceed, again for the sake of argument, on the assumption that we had the relevant Hilbert spaces. 

{\em Secundo}, the criterion, although perhaps necessary, is far removed from anything close to a sufficient condition, at least on its own. Consider the following three exhaustive possibilities. First, the physical Hilbert spaces of quantum geometrodynamics and LQG are both separable, i.e.\ they each have a countable basis. Second, one of them is separable, but the other is not. And third, both Hilbert spaces are non-separable, with either (a) their bases having the same cardinality, or (b) different cardinality. 

In the first case, the criterion is trivially satisfied because two bases in {\em any} two (infinite-dimensional) separable Hilbert spaces are unitarily equivalent. In the second case, the criterion is trivially violated, for corresponding reasons. In the third case, if the bases of the two Hilbert spaces have the same cardinality, we are back to the first situation; if they do not, we find ourselves in the second case again. So either way, the criterion by itself is not very illuminating and clearly not sufficient. It would have much more bite---and that may be the unarticulated intention behind Callender and Huggett's proposal---if it were augmented by some additional condition such as the preservation of the characteristic algebraic relations among the operators (such as the canonical commutation relations) in the transformation from one to the other. 

{\em Tertio}, the Callender-Huggett criterion gives the metric codification, which is used in quantum geometrodynamics, undue precedence over the connection codification, which is LQG's vantage point, in that it assumes that only the first captures the geometric essence of relativistic spacetimes. At least classically, both the metric and the connection descriptions are equally respectable ways of capturing the geometry of a spacetime and I see no reason to elevate one at the expense of the other. So we might, with equal justification, demand that a quantum theory of gravity offers a description of a quantum {\em spacetime} just in case a basis of its physical Hilbert space is unitarily equivalent to the connection basis of the physical Hilbert space of a quantum theory of gravity based on a connection representation.\footnote{Strictly speaking, LQG basic variables are the holonomies and fluxes introduced in \S\ref{ssec:lqg}, which are not identical to the connection and the canonically conjugate electric field of the connection representation but are constructed from them.} Such a choice would be, of course, vulnerable to the same charge raised here.

Thus, unitary equivalence between a basis of the physical Hilbert space of a theory in question and the three-metrics basis of quantum geometrodynamics is certainly not sufficient to think that the fundamental structure proposed by the theory in question is still spacetime. Perhaps it is not even necessary. But even if the criterion were valuable, we would still be faced with a rather complete dissolution of the classical continuous and local spacetime structure into granular structure with odd non-localities, represented by labelled graphs. And the question would still naturally arise how come our world looks like it is well described at sufficiently large scales by relativistic spacetimes. This explanation would still be owed, even if we managed to convince ourselves that the fundamental structure still deserves to be called `spacetime'.

\section{Re-emergence of spacetime}\label{sec:reemerge}

Before we venture into the enterprise of investigating how spacetime emerges from spin networks, one mistaken argument should be put to the side. I am thinking of a Kantian who nonchalantly responds to the present situation of the fundamental loss of spacetime by declaring that spacetime is a `pure form of intuition' and as such does not exist mind-independently anyway. So, the Kantian continues, we should not have expected to find spacetime as an ontological posit of a fundamental theory in the first place. But such a complacent `told-you-so' reaction would be entirely misguided; assuming space and time to be pure forms of intuition does nothing to relieve us from the obligation to explicate how relativistic spacetimes emerge from what physics tells us is fundamental. On a Kantian perspective, the job of physics is to describe nature as it appears to us, not as it may be {\em an sich}. And the natural world surely appears to be spatiotemporally ordered, which is why (earlier) physical theories made the natural assumption that there are space and time. Since physical theories involving such postulations have been empirically very successful, any theory seeking to supplant a theory as successful as, e.g., GR, must explain why the latter was as successful as it was given that it is not true. In this sense, recovering spacetime from the fundamental structure becomes part of the task of justifying the fundamental theory. This aspect assumes great urgency in a field plagued by the lack of empirical data. 

\begin{figure}
\centering
\epsfig{figure=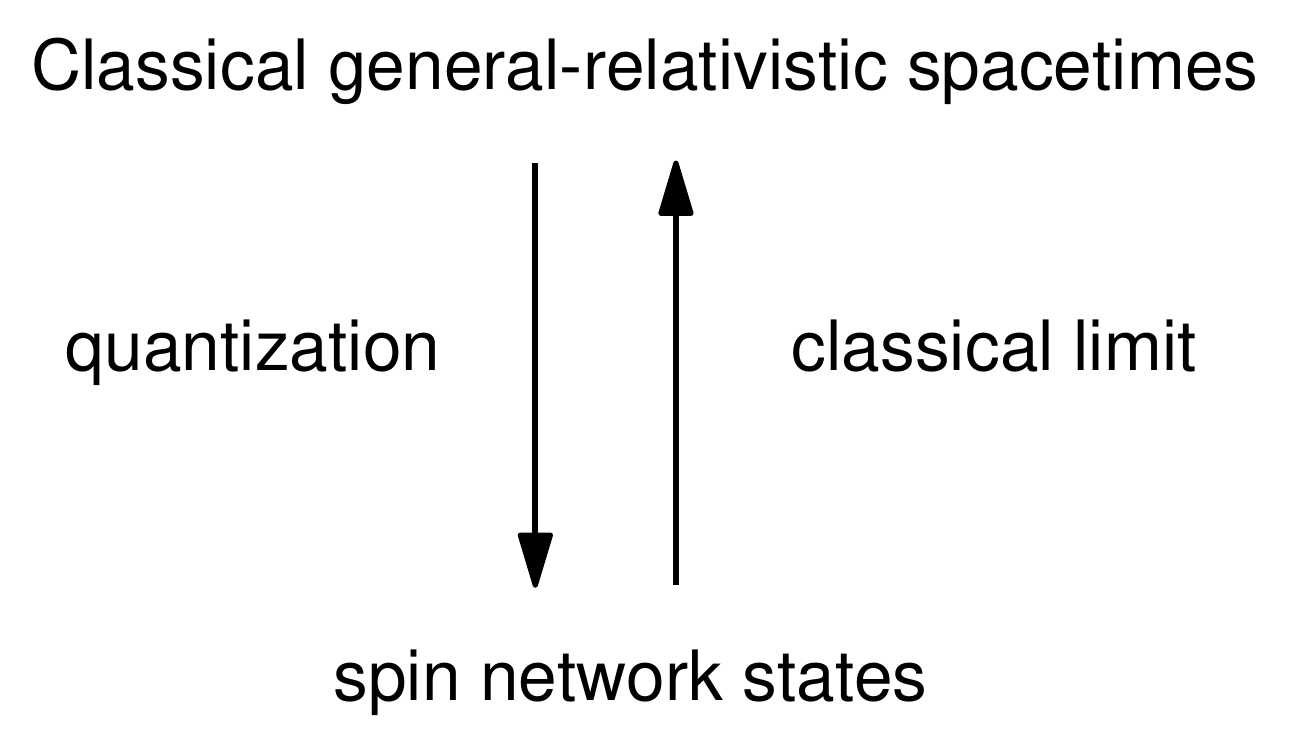,scale=0.55}
\caption{\label{fig:justification} Quantization and the classical limit as `inverse' tasks.}
\end{figure}

This justificatory task of understanding the emergence of spacetimes from fundamental structures such as spin networks is discharged by `taking the classical limit' of the fundamental theory: one shows that the classical theory results from an appropriate mathematical procedure which is interpreted to physically explain why and how the proprietary effects of the fundamental theory are hidden behind the phenomena so well represented by the classical theory. To express the situation in Reichenbachian terms, taking the classical limit, and thus showing how relativistic spacetimes emerge from fundamental structures, constitutes, at least partially, the `context of justification'. As indicated in Figure \ref{fig:justification}, the reverse process by which we arrived at the quantum theory of gravity from the classical theory is of course the quantization studied in \S\ref{sec:dissolve} and can thus be understood as the `context of discovery' (of the novel quantum theory). Understanding how classical spacetimes re-emerge is thus not only important to save the appearances and to accommodate common sense, but also a methodologically central part of the entire enterprise of quantum gravity. 

Nota bene, the quantization procedure as outlined above lacks a unique implementation for which every step is well justified. At various steps, one can choose to follow different paths, all presumably leading to inequivalent quantum theories. Some may find the fact that the construction of the quantum theory does not proceed along more principled lines troublesome. Applying this Reichenbachian terminology also illustrates why this need not be a problem: the `context of discovery' is dominated by creative elements which defy being bound by the narrow strictures of a research logic. On the other hand, the same traditional philosophy of science also urges that the other direction, the `context of justification', be taken very seriously. Regardless of this traditional philosophy of science's merits, the urgency clearly applies to the case at hand.\footnote{The remainder of this section draws on \citet[\S9]{wut06}.}

`Taking the classical limit' means establishing a mapping between, in some principled way, either individual models of the fundamental, `reducing', theory to individual models of the higher-level, `reduced', theory, or `generic' models of the reducing theory to `generic' models of the reduced theory, or the totality or near-totality of models of the reducing theory to the totality or near-totality of models of the reduced theory. It will not suffice to just procure a merely mathematical expression of such a mapping; instead, any formal articulation of it will need to be supplemented by a demonstration of its `physical salience' \citep{hugwut13b}. To start with the obvious, the map from the set of quantum states to the set of classical spacetimes should not be expected to be bijective, but many-to-one as there will be multiple distinct quantum states with the same classical limit.\footnote{Consider the $n$-body problem: while the phase space of states of an $n$-particle system in a physical space of $m$ dimensions is topologically $\mathbb{R}^{2mn}$ and therefore finite-dimensional in classical mechanics, the corresponding quantum space of states is the infinite-dimensional Hilbert space $L^2(\mathbb{R}^{mn})$, the space of square-integrable functions on $\mathbb{R}^{mn}$.} Furthermore, there will be no classical analogue for some sets of quantum states. Also, the quantization of a classical theory might not guarantee the re-emergence of the classical structure from the resulting quantum theory, due to interpretational issues \citep[80]{butish01}. 

So far, the classical limit of LQG (and many other quantum theories of gravity) has resisted understanding. The difficulties tend to be of two disparate kinds. First, there are technical intricacies. Secondly, and of present interest, there are numerous conceptual and interpretational issues. This is where philosophers can hope to make contributions by helping to explore the conceptual landscape, to map possibilities, and, more concretely, bring the literature on emergence and reduction to bear on the problems at hand. To date, only few philosophers have ventured into this area. I hope that more will follow---and there are hopeful signs. But still, \citet{butish99} and \citet{butish01} constitute more or less the complete philosophical literature on emergence in canonical quantum gravity, together with my dissertation \citep[\S9.2]{wut06}, on which the remainder of this section is based. 

A caveat before we proceed to portray the emergence scheme proposed by Jeremy Butterfield and Chris Isham and articulate its application to LQG and hence to the emergence of the full spacetime, rather than just time, as Butterfield and Isham do. As we noticed above (in \S\ref{ssec:lqg}), LQG is not a complete theory in that the `dynamics' is not well understood and in this sense the physical Hilbert space has not yet been isolated. Therefore, what follows below is limited to the kinematical level. This has some of the advantages of theft over honest toil, as we can thus circumvent the notorious problem of time, which of course Butterfield and Isham address. But it brings with it the distinct disadvantage that the following remains preliminary and must thus be taken with a grain of salt.

\subsection{The Butterfield-Isham scheme}\label{ssec:buttishscheme}

Let us then orient our conceptualization of the problem toward the extant literature on emergence in canonical quantum gravity. Similarly to my suggestion above, \citet{butish99,butish01} propose to regard quantization and emergence as two distinct, somewhat inverse, and {\em independent} strategies for solving the problem of quantum gravity. Butterfield and Isham consider various potentially helpful explications of the concept of emergence. As it turns out, all of them cast emergence as a reductive relation. As we have seen in \S\ref{ssec:nonred}, this usage is consonant with the physics literature, but dissonant with the one in general philosophy. Given the richness and diversity of the literature on reductive relations between theories, \citet{butish99} conclude that this should be taken to sustain the conclusion that there may not be a single concept of reduction to fit all instances considered, not even if the analysis is confined to physics.\footnote{No attempt shall be made to substantially consider the wider literature on the topic. Cf.\ \citet{spe78} for an analysis of various proposals for reduction as an inter-theoretic relation, with a particular eye on the physical sciences.} 

\citet{butish99} distinguish three ways in which theories (or their concepts, entities, laws, or models) can stand in a reductive relation to one another: definitional extension, supervenience, and emergence. The first typically assumes a syntactic understanding of theories, i.e.\ it understands a theory as a deductively closed set of propositions. Applying Butterfield and Isham's definition of it to the case at hand, one could say that GR is a {\em definitional extension} of LQG iff it is possible to add to LQG definitions of all non-logical symbols of GR such that every theorem of GR can be proven in LQG thus augmented. The concept of definitional extension is attractive because it gives us a clear understanding of how two theories, one of which is a definitional extension of the other, relate to one another. Thus, definitional extension goes a long way to explain why the predecessor theory was as successful as it was and why it breaks down where in fact it does. However, we do not expect the relation between GR and LQG to be as clear-cut as it is between Newtonian mechanics and special relativity, where the concept of definitional extension admits a rather straightforward application. In order to determine whether or not GR is a definitional extension of LQG, one would need to know how to recover the classical limit. Unless there is at least some progress in the recovery of the classical limit of LQG, the concept of definitional extension cannot usefully be applied to the case at stake. One would expect, to be sure, that relating LQG to GR will involve approximations such that general-relativistic propositions only hold approximately in LQG, and only under certain conditions. More specifically, one first extends the definitions of LQG such as to make it conceptually sufficiently potent to be able to prove all theorems of an intermediate theory, from which GR can, in a well-understood way, be recovered as an approximation. This process of approximation can be defined as follows:
\begin{defi}[Approximating procedure]\label{def:approx}
An {\em approximating procedure} designates the process of either neglecting some physical magnitudes, and justifying such neglect, or selecting a proper subset of states in the state space of the approximating theory, and justifying such selection, or both, in order to arrive at a theory whose values of physical quantities remain sufficiently close to those of appropriately related quantities in the theory to be approximated. 
\end{defi}
But all of this goes beyond the concept of definitional extension and shall be discussed below when I will discuss approximation as a form of emergence.\footnote{The clause ``appropriately related quantities in the theory to be approximated'' in Definition \ref{def:approx} above occludes substantive work that must be completed to achieve such ``appropriate relation''. I am grateful to Erik Curiel for pushing me on this point---I most certainly deserve the pushing here.}

The second relation considered by Butterfield and Isham is supervenience. {\em Per definitionem}, GR {\em supervenes} on LQG iff all its predicates supervene on the predicates of LQG, with respect to a fixed set $\mathfrak{A}$ of objects on which both predicates of GR and of LQG are defined. The set of predicates of GR is said to {\em supervene} on the set of predicates in LQG, given a set $\mathfrak{A}$ of objects, iff any two objects in $\mathfrak{A}$ that differ in what is predicated of them in GR must also differ in what is predicated of them in LQG. The fact that supervenience requires a stable set $\mathfrak{A}$ of objects underlying both theories, i.e.\ an identical ontology on which the ideologies of both theories are defined, renders it rather useless in the present case. In a very rough way, the ontology of both theories of course contains the gravitational field. But the finer structure of the ontologies of both theories do not resemble each other: in LQG, one might perhaps find loops, or spin networks, or more generally the inhabitants of the physical Hilbert space in its ontology, while in GR, no such objects can be found. Hence, supervenience, at least as defined above, does not offer any help in understanding the relation between GR and LQG. Of course, the requirement that the set $\mathfrak{A}$ must underlie both theories can be relaxed: one could instead demand that the set $\mathfrak{A}$ of objects on which the sets $\mathfrak{P}_1$ and $\mathfrak{P}_2$ of properties figuring in the two theories are defined must be closed under compositional operations such as mereological sums or the formation of sets. The sets $\mathfrak{P}_1$ and $\mathfrak{P}_2$ would then be defined with respect to some base individuals, forming subsets $\mathfrak{A}_1$ and $\mathfrak{A}_2$ of $\mathfrak{A}$. Typically, these predications would induce some properties on the non-basic composite objects. Conceivably, this relaxation might be sufficient to overcome the disjointness of the sets $\mathfrak{A}_1$ and $\mathfrak{A}_2$.\footnote{I wish to thank Jeremy Butterfield for suggesting this relaxation.}

Consequently, we should not harbour any hope that GR either is a definitional extension of LQG or supervenes on LQG. However, if one admits a sufficiently liberal notion of emergence, hope resurges. The third broadly reductive relation proposed by Butterfield and Isham, and termed `emergence' by them, fits the bill:
\begin{defi}[Emergence]
For Butterfield and Isham, a theory $T_1$ {\em emerges} from another theory $T_2$ iff there exists either a limiting or an approximating procedure to relate the two theories (or a combination of the two).
\end{defi}
The definition of `approximating procedure' was given in Definition \ref{def:approx}; here is the one for `limiting procedure':
\begin{defi}[Limiting procedure]
A {\em limiting procedure} is taking the mathematical limit of some physically relevant parameters, in general in a particular order, of the underlying theory in order to arrive at the emergent theory.
\end{defi}
For it to have any prayer of sufficing to relate two theories, a limiting procedure as envisioned by Butterfield and Isham must be accompanied by a specification of a map between the theories that relates at least some of their algebraic or geometric structures.\footnote{Thanks to Erik Curiel for holding me to task here.} For both technical and conceptual reasons, one should not expect that the emergence of GR from LQG can be understood only as a simple limiting procedure. Carlo \citet[\S6.7.1]{rov04} delivers an account of how limiting procedures alone are incapable of establishing the missing link. He relates how loop quantum gravitists have not suspected that quantum space might turn out to have a discrete structure during the period from the discovery of the loop representation of GR around 1988 to the derivation of the spectra of the area and volume operators in 1995. He reminisces how during this period researchers believed that the classical, macroscopic geometry could be gained by taking the limit of a vanishing lattice constant of the lattice of loops. This limiting procedure was taken to run analogously to letting the lattice constant of a lattice field theory go to zero and thus define a conventional QFT. With this model in mind, something remarkable happened when people tried to construct so-called weave states which are characterized as approximating a classical metric: when the quantum states were defined as the limit one gains when the spatial loop density grows to infinity, i.e.\ when the loop size is assumed to go to zero, it turned out that the approximation did not become increasingly accurate as the limit was approached. This can be taken as a clear indication that taking this limit was physically inappropriate. What was observed instead was that eigenvalues of the area and volume operators increased. This, of course, meant that the areas and volumes of the spatial regions under consideration also increased. In other words, the physical density of the loops did not increase when the `lattice constant' was decreased. The physical density of loops, it turned out, remains unaffected by how large the lattice constant is chosen; it is simply given by a dimensional constant of the theory itself, Planck's constant.  This result is interpreted to mean that there is a minimal physical scale. Or, in Rovelli's words, ``more loops give more size, not a better approximation to a given [classical] geometry.'' (ibid.) The loops, it turns out, have an intrinsic physical size. Taking this limit, then, does not change the structure from discrete quantum states to smooth manifolds. It just does not change anything in the physics, except that we look at larger volumes. As some of the features of the classical geometry such as smoothness cannot be reduced to or identified with properties of the quantum states of the more fundamental theory, GR in toto does not reduce to LQG. Thus, a limiting procedure, at least if used in isolation, will just not do the trick.

On the conceptual side, a limiting procedure never eliminates superposition states, which of course are generic in a quantum theory. For this reason alone, a limiting procedure cannot succeed in recovering a classical theory from a quantum one. As argued by Klaas \citet{lan06}, the classical world only emerges from the quantum theory if some quantum states and some observables of the quantum theory are neglected, {\em and} some limiting procedure is executed. According to his view, to be discussed below, relating the classical with the quantum world thus takes both, the limiting as well as the approximating, procedures. 

Turning to approximations then, a series of theories the last of which will mimic classical spacetimes via approximations needs to be constructed. First, let us consider what the `approximandum', the classical theory to be approximated, should be. In GR, and in quantum theories based on it, one standardly, and perhaps somewhat unprincipledly, distinguishes between gravity and matter---a distinction routinely downplayed in particle-physics based approaches. They differ in their role and where they show up in the Einstein equations: gravity, the ``marble'' as Einstein called it, constitutes the left-hand side of the equations and determines the spacetime geometry; matter, the ``low-grade wood'', enters the stress-energy tensor on the right-hand side. In the quantization that led to LQG, no matter was assumed to be present: LQG results from a vacuum quantization of GR. It would seem, therefore, that states in LQG's physical Hilbert space should generically give rise to semi-classical states which yield emergent classical spacetimes that are vacuum solutions. But this expectation may be disappointed, and perhaps for a reason: it has been claimed that matter is implicitly built into LQG and that it would therefore be a mistake to think that no matter is present in spin network states. In particular, it may be that the very structure of the spin networks gives rise to matter in the appropriate low-energy limit. This means that it may be advisable not to be fixated on vacuum spacetimes. 

Similarly, Hamiltonian GR is restricted to spacetime models with topology $\Sigma\times\mathbb{R}$. Should we thus expect that the procedure for recovering relativistic spacetimes would only yield spacetimes of such topology? While spacetimes with different topology may be suppressed and the generic result thus be concentrated on $(\Sigma\times\mathbb{R})$-spacetimes, the quantum structures with their combinatorial and topologically variegated connections may lead to spacetimes with more complicated topologies than those permitted by Hamiltonian GR. 

In order to prepare the field for applying the Butterfield-Isham scheme, let us consider the major ways in which classical physics is typically held to relate to quantum physics, as listed and discussed, e.g., by \citet{lan06}: (i) by a limiting procedure involving the limit $\hbar\rightarrow 0$ for a finite system, (ii) by a limiting procedure involving the limit $N\rightarrow \infty$ of a large system of $N$ degrees of freedom while $\hbar$ is held constant, and (iii) either by decoherence or by a consistent histories approach. Landsman defends the point of view that while none of these manners is individually sufficient to understand how classicality emerges from the quantum world, they jointly suggest that it results from ignoring certain states and certain observables from the quantum theory.\footnote{For a more thorough discussion of Landsman's argument, cf.\ \citet[\S9.2.1]{wut06}.} 

As Landsman shows, taking limits such as $N\rightarrow \infty$, albeit `factual', i.e., pertaining to our world, and hence physically more reasonable, is mathematically just a special case of the `counterfactual', and hence physically more problematic, limit $\hbar\rightarrow 0$. Regardless of their physical salience, these limits will in themselves not suffice because no such limit can ever resolve a quantum superposition state into a classical state. Thus, something more will be necessary, and that is where many think `decoherence' will come into play. The main idea of the program of decoherence is that the generically assumed presence of interference in quantum states is suppressed by the system's interaction with the `environment', such as is thought to occur in the measurement process.\footnote{For reviews of decoherence, see \citet{bac12} and \citet{sch04}.} Decoherence, then, is the phenomenon that pure quantum states, by virtue of their interaction with the system's `environment', evolve, over very short time spans, from superposition states to `almost' mixed states with classical probability distributions but `almost' no quantum interference left. Roughly speaking, decoherence leaves the quantum system, to a high approximation, in an eigenstate of a macroscopically relevant operator; the classical probabilities of the resulting mixed states then only reflect our ignorance as to {\em which} eigenstate the system's in. 

Given that the system at stake is the universe, and all of it, of course, the notion that `environmental' degrees of freedom are those which decohere the system must be generalized so as to include `internal' degrees of freedom of the system. This does not mean that the system is put in a mixed state from the beginning---that would be begging the question, as a referee correctly remarked---, but instead to `coarse grain' and thereby `wash out' many degrees of freedom, which then effectively act as the environment of the `system' consisting of the remaining, physically salient degrees of freedom. This `internal' environment then induces the decoherence of the originally pure state. We will return to this `cosmological' problem below in the specific context of LQG.

The cosmological problem thus requires that we operate with a generalized notion of decoherence, which does not rely on a decohering system being embedded in an environment which is literally external to it. There is, however, a second issue that needs to be addressed. Decoherence is usually understood as a {\em dynamical} process of a system interacting with a large number of `environmental' degrees of freedom. How should we conceive of a dynamical process in the general quantum-gravitational context in which time itself is part of the system at stake and, at least for canonical approaches, in which we face the nasty problem of time? Unfortunately, I have no solution to offer here, but can only note the puzzling problem and venture a guess as to the direction in which its resolution may have to go. In my view, the solution will come from a considered understanding of how dynamical processes such as decoherence can co-emerge with spacetime such that the emergence of the former facilitates the emergence of the latter, {\em and vice versa}, to let dynamics and spacetime mutually enable one another. 

In sum, {\em if}---and only if---a theory of decoherence manages to give us a handle on how to identify the relevant degrees of freedom, and under what circumstances the interaction between these degrees of freedom and those which were not picked out as `environmental' leads to a suppression of interference, and how this suppression works in detail, particularly concerning its `dynamics', {\em then} we will have a mechanism that `drives the system' to the right sorts of semi-classical quantum states. In other words, such a mechanism would then justify the selection of the subset of states (and of a subset of physical magnitudes) we made in what Butterfield and Isham called the `approximating procedure'. 

\begin{figure}
\centering
\epsfig{figure=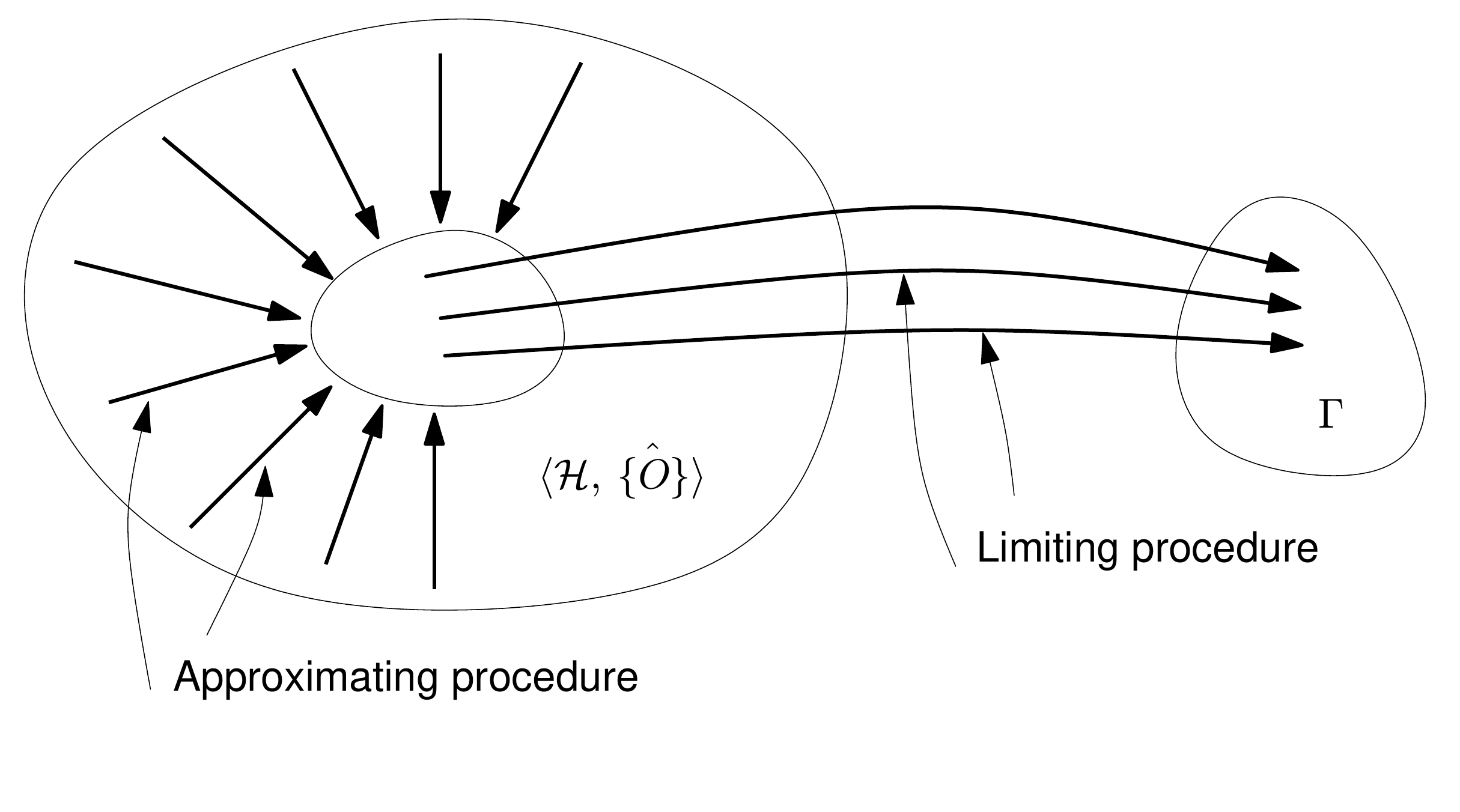,scale=0.55}
\caption{\label{fig:buttisham} The Butterfield-Isham scheme transposed to the present case.}
\end{figure}

In general outline, then, following Butterfield and Isham's proposal will lead to a two-step procedure, as illustrated in Figure \ref{fig:buttisham}. The first step consists of an approximating procedure, driving the generic quantum states, by some physical mechanism or other, into the semi-classical states, which are more closely related to classical states. The second step involves a limiting procedure relating these semi-classical states to states in the classical phase space, denoted in Figure \ref{fig:buttisham} by $\Gamma$. Regardless of how the details of this story work out, one thing is clear: a whole host of issues known from the traditional problem of understanding the relation between the quantum and the classical world will arise.

\subsection{Applying it the Butterfield-Isham scheme}\label{ssec:applying}

The thesis---or should I say the `promissory note'---to be suggested in the remainder of this essay asserts that at least to the extent to which LQG is a consistent and complete theory, (a close cousin of) GR can be seen to emerge from LQG if a delicately chosen ordered combination of approximations and limiting procedures is applied. This note is yet to be redeemed. All approaches to finding the semi-classical and classical limits of LQG are confined, to date, to using the kinematical Hilbert space $\mathcal{H}_K$ rather than the physical Hilbert space $\mathcal{H}$ as their starting point. This raises the concern of both the viability and the meaningfulness of relating the kinematical states to corresponding classical spacetimes, or spaces. But concerns like these, although perhaps ultimately critical, should not keep us from attempting to get a grasp on what it means to draw the classical limit of the background-independent QFT as it stands now (and has been sketched above), as it may turn out to be an eminent help in the construction of the physical Hilbert space itself. To be sure, even the relationship between kinematic LQG and classical theories is ill-understood. Let me sketch, however, how preliminary work by physicists might bear out the Butterfield-Isham scheme. 

The rough idea of constructing semi-classical states from the kinematical Hilbert space $\mathcal{H}_K$ is to find those kinematical states which correspond to almost flat three-metrics, i.e.\ to three-geometries where the quantum fluctuations are believed to be negligibly small. Two major approaches to construct semi-classical theories dominate the extant literature, the so-called `weave state approach' and the ansatz using coherent states. The latter has been pioneered by Thomas Thiemann and Oliver Winkler.\footnote{For a review, cf.\ \citet{saheal01} and \citet[\S11.2]{thi07}. Thiemann's book also discusses weave states in \S11.1 and the photon Fock states in \S11.3.} Other proposals include Madvahan Varandarajan's `photon Fock states' and generalizations thereof \citep{ashlew01,van00}, and the Ashtekar group's `shadow states' \citep{asheal03b}.\footnote{As \citet[\S11]{thi07} points out, there are deep connections between the various semi-classical programmes.} The remainder of this essay shall be dedicated, however, to the most prominent approach of constructing semi-classical states, the so-called `weave states'.

The idea of a {\em weave state} originally introduced by \citet*{ashrovsmo92},\footnote{For an intuitive introduction, see \citet[\S6.7.1]{rov04}. The picture is that of the gravitational field like a (quantum cloud of) fabric(s) of weaves which appears to be smooth if seen from far but displays a discrete structure if examined more closely. Hence {\em weave states}.} revolves around selecting spin network states that are eigenstates of the geometrical operators for the volume of a (spatial) region $\mathcal{R}$ with eigenvalues which approximate the corresponding classical values for the volume of $\mathcal{R}$ as determined by the classical gravitational field. Simultaneously, these selected spin network states are eigenstates of the geometrical area operator for a surface $\mathcal{S}$. More technically, consider a macroscopic three-dimensional region $\mathcal{R}$ of spacetime with the two-dimensional surface $\mathcal{S}$ and the three-dimensional gravitational field $e^i_a(\vec{x})$ defined for all $\vec{x}\in \mathcal{R}$. This gravitational field defines a metric field $q_{ab}(\vec{x}) = e^i_a (\vec{x}) e^j_b (\vec{x}) \eta_{ij}(\vec{x})$, where $\eta_{ab}$ is the Minkowski metric, for which it is possible to construct a spin network state $|S\rangle$ such that $|S\rangle$ approximates the metric $q_{ab}$ for sufficiently large scales $\Delta \gg \pl$, where $\pl$ is the Planck length, in a yet to be rigorously specified sense.\footnote{\label{fn:prekinwarning}The `upper case' spin network states $|S\rangle$ live in $\mathcal{K}^\star$, the {\em pre-kinematical Hilbert space}, i.e.\ the Hilbert space containing all spin network states which solve the Gauss constraints, but not necessarily the spatial diffeomorphism constraints. Thus, the spin network states in $\mathcal{K}^\star$ are not represented by abstract graphs, as are those in the full kinematical Hilbert space $\mathcal{H}_K$, but as embedded graphs on a background manifold. This choice is just conveniently following the established standard in the literature on weave states; we will see below in Footnote \ref{fn:prekin2} that this poses no problem as everything can be directly carried over to the spatially diffeomorphically invariant level.} Classically, the area of a two-dimensional surface $\mathcal{S}\subset \mathcal{M}$ and the volume of a three-dimensional region $\mathcal{R}\subset \mathcal{M}$ with respect to a (sufficiently well-behaved) fiducial gravitational field $^0e^i_a$ are given by \citep[\S2.1.4]{rov04}
\begin{eqnarray}\label{eq:classarea}
\mathbf{A}[^0e,\mathcal{S}] &=& \int |d^2\mathcal{S}|,\\ \label{eq:classvol}
\mathbf{V}[^0e,\mathcal{R}] &=& \int |d^3\mathcal{R}|,
\end{eqnarray}
where the relevant measures for the integrals are determined by $^0e^i_a$. This fiducial metric is typically, but not necessarily, chosen to be flat. The requirement that the spin network state $|S\rangle$ must approximate the classical geometry for sufficiently large scales is made precise by demanding that $|S\rangle$ be a simultaneous eigenstate of the area operator $\hat{\mathbf{A}}$ and the volume operator $\hat{\mathbf{V}}$ as mentioned above with eigenvalues equal to the classical values as given by (\ref{eq:classarea}) and (\ref{eq:classvol}), respectively, up to small corrections of the order of $\pl/\Delta$:
\begin{eqnarray}\label{eq:weavearea}
\hat{\mathbf{A}}(\mathcal{S}) |S\rangle &=& \left( \mathbf{A}[^0e,\mathcal{S}] + \mathcal{O}(\pl^2/\Delta^2) \right) |S\rangle,\\ \label{eq:weavevol}
\hat{\mathbf{V}}(\mathcal{R}) |S\rangle &=& \left( \mathbf{V}[^0e,\mathcal{R}] + \mathcal{O}(\pl^3/\Delta^3) \right) |S\rangle.
\end{eqnarray}
If a spin network state $|S\rangle$ satisfies these requirements, then it is called a {\em weave state}. In fact, the length scale $\Delta$, which is large compared to the Planck length $\pl$, characterizes the weave states, which are for this reason sometimes denoted by $|\Delta\rangle$ in the literature. At scales much smaller than $\Delta$, the quantum features of spacetime would become relevant, while at scales of order $\Delta$ or larger, the weave states exhibit a close approximation to the corresponding classical geometry in the sense that it determines the same areas and volumes as the classical metric $q_{ab}$. In this sense, the weave states are semi-classical approximations.

It should be noted that the correspondence between weave states and classical spacetimes is many-to-one. In other words, equations (\ref{eq:weavearea}) and (\ref{eq:weavevol}) do not determine the state $|S\rangle$ uniquely from a given three-metric $q_{ab}$. The reason for this is that these equations only put constraints on values averaged over all of $\mathcal{S}$ and $\mathcal{R}$, respectively, and we have assumed {\em ex constructione} that these regions are large compared to the Planck scale. Of course, there are many spin network states with these averaged properties, but only one classical metric which exactly corresponds to these averages values. The situation can be thought of as somewhat analogous to thermodynamics, where a physical system with many microscopic degrees of freedom has many different microscopic states with the same averaged, macroscopic properties such as temperature.\footnote{\label{fn:prekin2}The weave states as introduced above have merely been defined at the pre-kinematic level, i.e.\ they are not formulated in terms invariant under spatial diffeomorphisms (cf.\ also Footnote \ref{fn:prekinwarning}). The reason for this choice lies mostly in that this is the canonical choice in the literature, but also because in this way, the weave states can be directly related to three-metrics, rather than equivalence classes of three-metrics. This, however, does not constitute a problem whatsoever, as the characterization of weave states carries over into the context of diffeomorphically invariant spin network states in $\mathcal{H}_K$, as follows. If we introduce a map $P_{\mbox{\scriptsize diff}}: \mathcal{K}^\star\rightarrow \mathcal{H}_K$ which projects states in $\mathcal{K}^\star$ related by a spatial diffeomorphism unto the same element of $\mathcal{H}_K$, then the state $\mathcal{H}_K\ni |s\rangle = P_{\mbox{\scriptsize diff}} |S\rangle$ is a {\em weave state} of the classical three-geometry $[q_{ab}]$, i.e., the equivalence class of three-metrics $q_{ab}$ under spatial diffeomorphisms, just in case $|S\rangle$ is a weave state of the classical three-metric $q_{ab}$ as defined above.}

Apart from a serious difficulty in constructing semi-classical weave states corresponding to classical Minkowski spacetime,\footnote{For details, cf.\ \citet[181]{wut06}.} it seems as if the notion of approximation as captured in Definition \ref{def:approx} and the Butterfield-Isham scheme might bear fruit in relating semi-classical weave states to classical spacetimes (or at least spaces). If the weave states are taken to be simultaneous eigenstates of the area and volume operators, as they are in (\ref{eq:weavearea}) and (\ref{eq:weavevol}), then some physical quantities must be neglected, viz.\ all those operators constructed from connection operators, since the `geometrical' eigenstates are maximally spread in these operators, and the kinematical (weave) states must be carefully selected to only include those which are peaked around the geometrical values determined by the fiducial metric. It is at least questionable, however, whether the neglect of connection-based operators can be justified. If it cannot, then only semi-classical states which are peaked in both the connection and the triad basis, and are peaked in such a manner as to approximate classical states, should be considered. In this case, we would still only have a selection of states, but perhaps no operators, or no physically salient ones, which are being ignored. 

None of this gives us just as yet a {\em physical mechanism} that drives generic kinematical states to the semi-classical weave states. Just as above in the general case, decoherence is widely assumed to offer such a mechanism in the context of weave states. But this brings what I termed above the `cosmological problem' back into the fold: how should such a story possibly apply to the present context where the spin network states are supposed to be the quantum account of space---and all of it. If we thus think of an `environment' as something external to the system for which it is an environment, then relying on such an environment in our story implies that there must be something outside of space. But this is clearly incoherent. Not all hope is lost, however, as there are at least two ways to escape the incoherence. First, as in the general case above, one might conceive of decoherence not in terms of external, environmental degrees of freedom which interact with the system, but instead as interactions among different degrees of freedom of the system itself. This will presuppose a partition of the system's degrees of freedom into `salient' ones and mere `background'; but there is no reason that this couldn't be done in a principled fashion. 

Secondly, we may reconceptualize LQG's subject matter. We may, more specifically, conceive of areas and volumes as local properties of the quantum gravitational field, just as these geometrical properties were local in GR. As was explicated in \S\ref{ssec:lqg}, given a region $\mathcal{R}$ of quantum space, e.g.\ a chunk of space in our laboratory, each node of the spin network state represents a grain of such a space as it contributes to the eigenvalue of the volume operator. Similarly, each link from a node within $\mathcal{R}$ to a node outside of $\mathcal{R}$, i.e.\ each link which intersects the boundary $\mathcal{S}$ of $\mathcal{R}$, contributes to the eigenvalue of the area operator. If we had measurement devices at our disposal with Planck-scale accuracy, we could, in principle, measure the volume and the surface area of a region of space(time) given in our lab. Such a measurement would essentially amount to counting (and weighing) the nodes within a region as well as counting (and weighing) the links which leave the region. If the region $\mathcal{R}$ considered does not encompass all of space, but only a delimited piece of it, then of course finding an environment for such a `mid-sized' region is straightforward and the cosmological problem dissolves. In fact, it would arguably also resolve the dynamical problem, as the lab frame would offer a context in which dynamical processes unravel. It could thus be the case that if we performed an area or volume measurement on surface $\mathcal{S}$ or region $\mathcal{R}$, respectively, then we would find the quantum state of this `mid-sized' region decohered into an eigenstate of the relevant operator, and thus into a weave state.

Once we have completed this stage, and we have found semi-classical states which approximate classical states, then a limiting procedure can be executed. Such a limiting procedure will involve taking the limit $\pl/\Delta\rightarrow 0$, which will make the small corrections in (\ref{eq:weavearea}) and (\ref{eq:weavevol}) disappear. This limit can be performed by either having $\Delta$ go to infinity, or $\pl$ go to zero (or both). The first choice corresponds to letting the size of the spatial region $\mathcal{R}$ grow beyond all limits, and thus resembles the `factual' limit $N\rightarrow \infty$ as discussed above. The second choice, letting the Planck size go to zero, corresponds, accordingly, to the `counterfactual' case $\hbar\rightarrow 0$. With the second choice, but arguably not the first, we leave the realm of the quantum theory and arrive at a strictly classical description of the spatial geometry.

It should be noted that none of this solves the measurement problem. Only a full solution of the measurement problem will ultimately give us complete comprehension of the emergence of classicality from a reality which is fundamentally quantum. But to solve this problem is hard in non-relativistic quantum mechanics, harder still if special relativity must be incorporated, and completely mystifying once we move to fully relativistic quantum theories of gravity. In light of this, I submit that we would have reason to uncork our champagne even if we only managed to articulate a complete and consistent quantum theory of gravity with a well-understood approximation to semi-classical states and a somewhat rigorous limiting procedure connecting these semi-classical states to classical states of the gravitational field.

\section{Conclusion}\label{sec:conc}

We have seen how classical space and time `disappear' in quantum gravity and considered a sketch of how they might re-emerge from the fundamental, not obviously spatiotemporal structure. Even though the situation is technically and conceptually more demanding overall and even though a {\em case} must be made for the applicability of a traditional measurement concept more specifically, I hope the reader has also recognized that the way in which classicality emerges from the quantum theory does not radically differ from ordinary quantum mechanics, at least along some dimensions of comparison. 

The project of analyzing the emergence of spacetime, and hence of classicality, from quantum theories of gravity, which often deny at least some aspects of spatiotemporality, is relevant for two reasons. First, important foundational questions concerning the interpretation of, and the relation between, theories are addressed, which contributes to the conceptual clarifications in the foundations of physics arguably necessary to achieve a breakthrough. Not only philosophers of physics will contribute to this project, of course. They are not even likely to shoulder the lion's share, which will still fall on the physicists. But they can nevertheless bring their unique skill set to the table, to the benefit, it is hoped, of the entire dinner party. Secondly, and conversely, quantum gravity is rich with implications for specifically philosophical, and particularly metaphysical, issues concerning not just space and time, but also causation, reduction, and even modality. Quantum gravity thus turns out to be a very fertile ground for the philosopher. Altogether, I take it, there is no reason for philosophers to keep aloof from these exciting developments in the foundations of physics.

\bibliographystyle{plainnat}
\bibliography{/Users/christian/Professional/Bibliographies/quantumgravity}

\begin{thebibliography}{49}
\providecommand{\natexlab}[1]{#1}
\providecommand{\url}[1]{\texttt{#1}}
\expandafter\ifx\csname urlstyle\endcsname\relax
  \providecommand{\doi}[1]{doi: #1}\else
  \providecommand{\doi}{doi: \begingroup \urlstyle{rm}\Url}\fi

\bibitem[Ashtekar and Lewandowski(2001)]{ashlew01}
Abhay Ashtekar and Jerzy Lewandowski.
\newblock Relation between polymer and {Fock} excitations.
\newblock \emph{Classical and Quantum Gravity}, 18:\penalty0 L117--L128, 2001.

\bibitem[Ashtekar and Lewandowski(1997)]{ashlew97}
Abhay Ashtekar and Jerzy Lewandowski.
\newblock Quantum theory of geometry {I}: Area operators.
\newblock \emph{Classical and Quantum Gravity}, 14:\penalty0 A55--A81, 1997.

\bibitem[Ashtekar and Lewandowski(1998)]{ashlew98}
Abhay Ashtekar and Jerzy Lewandowski.
\newblock Quantum theory of geometry {II}: Volume operators.
\newblock \emph{Advances in Theoretical and Mathematical Physics}, 1:\penalty0
  388--429, 1998.

\bibitem[Ashtekar and Lewandowski(1999)]{ashlew99}
Abhay Ashtekar and Jerzy Lewandowski.
\newblock Quantum field theory of geometry.
\newblock In Tian~Yu Cao, editor, \emph{Conceptual Foundations of Quantum Field
  Theory}, pages 187--206, Cambridge, 1999. Cambridge University Press.

\bibitem[Ashtekar et~al.(1992)Ashtekar, Rovelli, and Smolin]{ashrovsmo92}
Abhay Ashtekar, Carlo Rovelli, and Lee Smolin.
\newblock Weaving a classical metric with quantum threads.
\newblock \emph{Physical Review Letters}, 69:\penalty0 237--240, 1992.

\bibitem[Ashtekar et~al.(2003)Ashtekar, Fairhurst, and Willis]{asheal03b}
Abhay Ashtekar, Stephen Fairhurst, and Joshua~L Willis.
\newblock Quantum gravity, shadow states, and quantum mechanics.
\newblock \emph{Classical and Quantum Gravity}, 20:\penalty0 1031--1062, 2003.

\bibitem[Bacciagaluppi(2012)]{bac12}
Guido Bacciagaluppi.
\newblock The role of decoherence in quantum theory.
\newblock In Edward~N. Zalta, editor, \emph{Stanford Encyclopedia of
  Philosophy}, 2012.
\newblock URL \url{http://plato.stanford.edu/entries/qm-decoherence/}.

\bibitem[Belot and Earman(2001)]{belear01}
Gordon Belot and John Earman.
\newblock {Pre-Socratic} quantum gravity.
\newblock In Craig Callender and Nick Huggett, editors, \emph{Physics Meets
  Philosophy at the Planck Scale}, pages 213--255. Cambridge University Press,
  Cambridge, 2001.

\bibitem[Butterfield and Isham(1999)]{butish99}
Jeremy Butterfield and Chris Isham.
\newblock On the emergence of time in quantum gravity.
\newblock In Jeremy Butterfield, editor, \emph{The Arguments of Time}, pages
  111--168. Oxford University Press, Oxford, 1999.

\bibitem[Butterfield and Isham(2001)]{butish01}
Jeremy Butterfield and Christopher Isham.
\newblock Spacetime and the philosophical challenge of quantum gravity.
\newblock In Craig Callender and Nick Huggett, editors, \emph{Physics Meets
  Philosophy at the Planck Scale}, pages 33--89. Cambridge University Press,
  Cambridge, 2001.

\bibitem[Callender and Huggett(2001)]{calhug01b}
Craig Callender and Nick Huggett.
\newblock Introduction.
\newblock In Craig Callender and Nick Huggett, editors, \emph{Physics Meets
  Philosophy at the Planck Scale}, pages 1--30. Cambridge University Press,
  Cambridge, 2001.

\bibitem[Choquet-Bruhat and York(1980)]{choyor80}
Yvonne Choquet-Bruhat and James~W York, Jr.
\newblock The {Cauchy} problem.
\newblock In Alan Held, editor, \emph{General Relativity and Gravitation: One
  Hundred Years After the Birth of Albert Einstein}, pages 99--172. Plenum
  Press, New York, 1980.

\bibitem[Curiel(2001)]{cur01}
Erik Curiel.
\newblock Against the excesses of quantum gravity: A plea for modesty.
\newblock \emph{Philosophy of Science}, 68\penalty0 (3):\penalty0 S424--S441,
  2001.

\bibitem[Curiel(2009)]{cur09}
Erik Curiel.
\newblock General relativity needs no interpretation.
\newblock \emph{Philosophy of Science}, 76\penalty0 (1):\penalty0 44--72, 2009.

\bibitem[Debnath and Mikusi\'nski(1999)]{debmik99}
Lokenath Debnath and Piotr Mikusi\'nski.
\newblock \emph{Introduction to Hilbert Spaces with Applications}.
\newblock Academic Press, San Diego, 1999.

\bibitem[Earman(2003)]{ear03}
John Earman.
\newblock Tracking down gauge: An ode to the constrained {H}amiltonian
  formalism.
\newblock In Katherine Brading and Elena Castellani, editors, \emph{Symmetries
  in Physics: Philosophical Reflections}, pages 140--162. Cambridge University
  Press, Cambridge, 2003.

\bibitem[Earman(2006)]{ear06a}
John Earman.
\newblock The implications of general covariance for the ontology and ideology
  of spacetime.
\newblock In Dennis Dieks, editor, \emph{The Ontology of Spacetime}, pages
  3--23. Elsevier, Amsterdam, 2006.

\bibitem[Earman(1995)]{ear95}
John Earman.
\newblock \emph{Bangs, Crunches, Whimpers, and Shrieks: Singularities and
  Acausalities in Relativistic Spacetimes}.
\newblock Oxford University Press, New York, 1995.

\bibitem[Friedrich and Rendall(2000)]{friren00}
Helmut Friedrich and Alan Rendall.
\newblock The {Cauchy} problem for the {Einstein} equations.
\newblock \emph{Lecture Notes in Physics}, 540:\penalty0 127--224, 2000.

\bibitem[Halmos(1951)]{hal51}
Paul~R Halmos.
\newblock \emph{Introduction to Hilbert Space and the Theory of Spectral
  Multiplicity}.
\newblock Chelsea Publishing Company, New York, 1951.

\bibitem[Henneaux and Teitelboim(1992)]{hentei}
Marc Henneaux and Claudio Teitelboim.
\newblock \emph{Quantization of Gauge Systems}.
\newblock Princeton University Press, Princeton, NJ, 1992.

\bibitem[Huggett and W\"uthrich(2013)]{hugwut13b}
Nick Huggett and Christian W\"uthrich.
\newblock Emergent spacetime and empirical (in)coherence.
\newblock \emph{Studies in the History and Philosophy of Modern Physics},
  44:\penalty0 276--285, 2013.

\bibitem[Huggett et~al.(2013)Huggett, Vistarini, and W\"uthrich]{hugeal13}
Nick Huggett, Tiziana Vistarini, and Christian W\"uthrich.
\newblock Time in quantum gravity.
\newblock In Adrian Bardon and Heather Dyke, editors, \emph{A Companion to the
  Philosophy of Time}, pages 242--261. Wiley-Blackwell, Chichester, 2013.

\bibitem[Kiefer(2012)]{kie12}
Claus Kiefer.
\newblock Quantum gravity: whence, whither?
\newblock In Felix Finster, Olaf M\"uller, Marc Nardmann, J\"urgen Tolksdorf,
  and Eberhard Zeidler, editors, \emph{Quantum Field Theory and Gravity:
  Conceptual and Mathematical Advances in the Search for a Unified Framework},
  pages 1--13. Springer, Basel, 2012.

\bibitem[Lam and Esfeld(2013)]{lamesf13}
Vincent Lam and Michael Esfeld.
\newblock A dilemma for the emergence of spacetime in canonical quantum
  gravity.
\newblock \emph{Studies in the History and Philosophy of Modern Physics},
  44:\penalty0 286--293, 2013.

\bibitem[Landsman(2006)]{lan06}
Nicolaas~P Landsman.
\newblock Between classical and quantum.
\newblock In Jeremy Butterfield and John Earman, editors, \emph{Handbook of the
  Philosophy of Science. Vol.\ 2: Philosophy of Physics}, pages 417--553.
  Elsevier B.V., Amsterdam, 2006.

\bibitem[Manchak(2011)]{man11b}
John~Byron Manchak.
\newblock What is a physically reasonable space-time?
\newblock \emph{Philosophy of Science}, 78:\penalty0 410--420, 2011.

\bibitem[Markopoulou and Smolin(2007)]{marsmo07}
Fotini Markopoulou and Lee Smolin.
\newblock Disordered locality in loop quantum gravity states.
\newblock \emph{Classical and Quantum Gravity}, 24:\penalty0 3813--3823, 2007.

\bibitem[Maudlin(2002)]{mau02}
Tim Maudlin.
\newblock Thoroughly muddled {McT}aggart: Or how to abuse gauge freedom to
  generate metaphysical monstrosities.
\newblock \emph{Philosophers' Imprint}, 2\penalty0 (4), 2002.

\bibitem[McLaughlin(1997)]{mcl97}
Brian McLaughlin.
\newblock Emergence and supervenience.
\newblock \emph{Intellectica}, 2:\penalty0 25--43, 1997.

\bibitem[O'Connor and Wong(2012)]{ocowon12}
Timothy O'Connor and Hong~Yu Wong.
\newblock Emergent properties.
\newblock In Edward~N. Zalta, editor, \emph{Stanford Encyclopedia of
  Philosophy}, 2012.
\newblock URL \url{http://plato.stanford.edu/entries/properties-emergent/}.

\bibitem[Prescod-Weinstein and Smolin(2009)]{presmo09}
Chanda Prescod-Weinstein and Lee Smolin.
\newblock Disordered locality as an explanation for the dark energy.
\newblock \emph{Physical Review D}, 80:\penalty0 063505, 2009.

\bibitem[Rovelli(2002)]{rov02}
Carlo Rovelli.
\newblock Partial observables.
\newblock \emph{Physical Review D}, 65:\penalty0 124013, 2002.

\bibitem[Rovelli(2004)]{rov04}
Carlo Rovelli.
\newblock \emph{Quantum Gravity}.
\newblock Cambridge University Press, Cambridge, 2004.

\bibitem[Rovelli(2011)]{rov11a}
Carlo Rovelli.
\newblock A new look at loop quantum gravity.
\newblock \emph{Classical and Quantum Gravity}, 28:\penalty0 114005, 2011.

\bibitem[Rovelli and Smolin(1995{\natexlab{a}})]{rovsmo95a}
Carlo Rovelli and Lee Smolin.
\newblock Discreteness of area and volume in quantum gravity.
\newblock \emph{Nuclear Physics B}, 442:\penalty0 593--622, 1995{\natexlab{a}}.
\newblock Erratum: {\em Nuclear Physics} B, {\bf 456}:734.

\bibitem[Rovelli and Smolin(1995{\natexlab{b}})]{rovsmo95b}
Carlo Rovelli and Lee Smolin.
\newblock Spin networks and quantum gravity.
\newblock \emph{Physical Review D}, 52:\penalty0 5743--5759,
  1995{\natexlab{b}}.

\bibitem[Ruetsche(2011)]{rue11}
Laura Ruetsche.
\newblock \emph{Interpreting Quantum Theories: The Art of the Possible}.
\newblock Oxford University Press, Oxford, 2011.

\bibitem[Sahlmann et~al.(2001)Sahlmann, Thiemann, and Winkler]{saheal01}
Hanno Sahlmann, Thomas Thiemann, and Oliver Winkler.
\newblock Coherent states for canonical quantum general relativity and the
  infinite tensor product extension.
\newblock \emph{Nuclear Physics B}, 606:\penalty0 401--440, 2001.

\bibitem[Schlosshauer(2004)]{sch04}
Maximilian Schlosshauer.
\newblock Decoherence, the measurement problem, and interpretations of quantum
  mechancis.
\newblock \emph{Reviews of Modern Physics}, 76:\penalty0 1267--1305, 2004.

\bibitem[Smeenk and W\"uthrich(2011)]{smewut11}
Chris Smeenk and Christian W\"uthrich.
\newblock Time travel and time machines.
\newblock In Craig Callender, editor, \emph{The Oxford Handbook of Philosophy
  of Time}, pages 577--630. Oxford University Press, Oxford, 2011.

\bibitem[Smolin(2009)]{smo09}
Lee Smolin.
\newblock Generic predictions of quantum theories of gravity.
\newblock In Daniele Oriti, editor, \emph{Approaches to Quantum Gravity: Toward
  a New Understanding of Space, Time and Matter}, pages 548--570. Cambridge
  University Press, Cambridge, 2009.

\bibitem[Spector(1978)]{spe78}
Marshall Spector.
\newblock \emph{Concepts of Reduction in Physical Science}.
\newblock Temple University Press, Philadelphia, 1978.

\bibitem[Thiemann(2007)]{thi07}
Thomas Thiemann.
\newblock \emph{Modern Canonical Quantum General Relativity}.
\newblock Cambridge University Press, Cambridge, 2007.

\bibitem[Varadarajan(2000)]{van00}
Madvahan Varadarajan.
\newblock Fock representations from {$U(1)$} holonomy algebras.
\newblock \emph{Physical Review D}, 61:\penalty0 104001, 2000.

\bibitem[Wald(1984)]{walgr}
Robert~M Wald.
\newblock \emph{General Relativity}.
\newblock The University of Chicago Press, Chicago, 1984.

\bibitem[W\"uthrich(2006)]{wut06}
Christian W\"uthrich.
\newblock \emph{Approaching the Planck Scale from a Generally Relativistic
  Points of View: A Philosophical Appraisal of Loop Quantum Gravity}.
\newblock PhD thesis, University of Pittsburgh, 2006.

\bibitem[W\"uthrich(2012)]{wut12b}
Christian W\"uthrich.
\newblock {Die allgemeine Relativit\"atstheorie als Ausgangspunkt einer
  Quantentheorie der Gravitation}.
\newblock In Michael Esfeld, editor, \emph{{Philosophie der Physik}}, pages
  306--324. Suhrkamp, Berlin, 2012.

\bibitem[W\"uthrich(2013)]{wut13a}
Christian W\"uthrich.
\newblock A la recherche de l'espace-temps perdu.
\newblock In Soazig Le~Bihan, editor, \emph{Pr\'ecis de philosophie de la
  physique}, pages 222--241. Vuibert, Paris, 2013.

\end{thebibliography}

\end{document}